\documentclass[graybox, nosecnum]{svmult}

\usepackage{amsmath,amssymb,bm,curves}

\usepackage{lineno}

\usepackage{mathptmx}       
\usepackage{helvet}         
\usepackage{courier}        
\usepackage{type1cm}        
\usepackage{makeidx}         
\usepackage{graphicx}        
\usepackage{multicol}        
\usepackage[bottom]{footmisc}
\usepackage{hyperref}        
\usepackage{soul}            

\makeindex             

\def\ubar{{\bar{u}}}
\def\dbar{{\bar{d}}}
\def\sbar{{\bar{s}}}

\begin{document}
\title*{The $\eta$- and $\eta'$-nucleus interactions and the search for $\eta$, $\eta'$- mesic states}
\author{Steven D. Bass, 
Volker Metag\thanks{corresponding author} and Pawel Moskal}
\institute{Steven Bass \at 
Kitzb\"uhel Centre for Physics, Kitzb\"uhel, Austria.
\at
Marian Smoluchowski Institute of Physics and Institute of Theoretical Physics, Jagiellonian University, Krak\'ow, Poland.\\
\email{Steven.Bass@cern.ch}
\and Volker Metag \at II. Physikalisches Institut, University of Giessen, 35392 Giessen, Germany.\\
\email{Volker.Metag@exp2.physik.uni-giessen.de}
\and Pawel Moskal \at Marian Smoluchowski Institute of Physics, Jagiellonian University, Krak\'ow, Poland.\\ 
\email{p.moskal@uj.edu.pl}}
%
%
\maketitle
\abstract{
The isoscalar $\eta$ and $\eta'$ mesons are special in QCD, being linked both to chiral symmetry and to non-perturbative glue associated with the axial anomaly.
The properties of these mesons in medium are sensitive to how these dynamics work in
the nuclear environment.
In contrast to pionic and kaonic atoms 
which are mainly bound by the Coulomb interaction with some corrections due to the strong force 
(Coulomb assisted binding), the $\eta$ and $\eta'$ as neutral mesons can only be bound by the strong interaction. Is this interaction strong enough? This topic has inspired a vigorous program of experiments, 
conducted in close contact with theory,
One has to determine the complex $\eta$, $\eta'$-nucleus potential. Does the real part $V$ provide a sufficiently deep potential? Is the imaginary part $W$ small enough to allow for narrow states that can more easily be detected experimentally, i.e. 
$|W| \ll |V|$?
The $\eta'$ effective mass is observed to be suppressed by 
$\approx -40$ MeV at nuclear matter density.
Bound state searches are ongoing.
This article gives
an overview of the status of knowledge in this field together with an outlook to future experiments.
\\
}

\section{\textit{1. Introduction}}

Low-energy Quantum Chromodynamics, QCD, is characterized
by 
confinement and dynamical chiral symmetry breaking.
The fundamental quark and gluon degrees of freedom
probed in high-energy deep inelastic scattering freeze out
into hadrons.
One finds nuclei built from protons and neutrons which interact through meson exchanges
(pions, $\rho$, $\omega$, ...)
together
with an important role for the $ {\rm \Delta}$ resonance in low-energy pion nucleon interactions.
In the absence of 
small quark mass contributions, 
the proton mass {\bf is} determined
by gluonic binding 
energy.
Pions and kaons are would-be Goldstone bosons associated with chiral symmetry
with mass{\bf es} squared proportional to the masses of their constituent quarks.
Besides the confinement
potential,
gluonic degrees are active in the flavour-singlet channel and increase by about 300-400 MeV the masses of their isosinglet partners, the $\eta$ and $\eta'$ states, which are then no longer pure Goldstone states.

Hadron properties are modified in nuclear media.
Studying these properties
using the nucleus
as a ``detector'' 
opens a new window on
low-energy QCD dynamics
including chiral symmetry.

One finds a small pion mass shift of order a few MeV in asymmetric nuclear matter 
\cite{Kienle:2004hq}.
Experiments with deeply bound 
pionic atoms reveal a reduction in 
the value of the pion decay constant
$f_{\pi}^{*2}/f_{\pi}^2 = 0.64 \pm 0.06$ 
at nuclear matter density \cite{Suzuki:2002ae}.
K$^-$-mesons are observed to experience an effective mass drop 
of the order of 200 MeV 
at about two times 
nuclear matter density in heavy-ion collisions 
\cite{Schroter:1994ck,Barth:1997mk}.
A detailed overview
of 
meson properties in 
medium is given in 
\cite{Metag:2017yuh}.
One also finds that the nucleon and ${\rm \Delta}$ masses are suppressed in medium 
\cite{Oset:1987re,Lenske:2018bgr,Lenske:2018bvq}.
What should one expect for the $\eta$ and $\eta'$?
How does the gluonic part of their mass behave in nuclei?
Can one find $\eta$ and $\eta'$ bound states in nuclei?
Without its gluonic mass contribution, the $\eta'$ would be a strange quark state with just small interaction with the light-quark meson mean fields present in the nucleus
\cite{Bass:2005hn,Bass:2013nya}.

Meson masses in nuclei are determined from the meson nucleus 
optical potential and the scalar induced contribution 
to the meson propagator evaluated at zero three-momentum, 
${\vec k} =0$, in the nuclear medium.
Let $k=(E,{\vec k})$ and $m$ denote the four-momentum and 
mass of the meson in free s
pace. 
Then, one solves the equation
\begin{equation}
k^2 - m^2 = {\tt Re} \ \Pi (E, {\vec k}, \rho)
\label{eq:eqa1}
\end{equation}
for ${\vec k}=0$
where $\Pi$ is the in-medium $s$-wave meson self-energy
and $\rho$ is the nuclear density.
Contributions to the in medium mass come from coupling to 
the scalar $\sigma$ field in the nucleus in mean field approximation, and to
nucleon-hole and resonance-hole excitations in the medium.
For ${\vec k}=0$, $k^2 - m^2 \sim 2 m (m^* - m)$ 
where $m^*$ is the effective mass in the medium.
The mass shift $m^*-m$ is the depth or real part of 
the meson nucleus optical potential. The imaginary part of 
the potential measures the width of the meson in the nuclear medium.

\newpage

The $s$-wave self-energy
can be written as 
\cite{Ericson:1988gk,Friedman:2007zza}
\begin{equation}
\Pi (E, {\vec k}, \rho) \bigg|_{\{{\vec k}=0\}}
=
- 4 \pi a \rho 
\biggl(
1 + 
\frac{A}{A-1}
\frac{\mu}{M} \biggr)
\label{eq:eqa2}
\end{equation}
Here 
$a$ is the meson-nucleon scattering length
and
$\rho$ is the nuclear density;
$A$ is the atomic number,
$M$ is the nucleon mass
and
$\mu = m M_A / (m+M_A)$
where $M_A$ is the mass of the nucleus.
The expression 
in
Eq.(\ref{eq:eqa2})
is quoted
to leading order in $a$;
that is, suppressing higher order terms in $a$ from 
Ericson-Ericson-Lorentz-Lorenz multiple scattering
corrections.

Attraction corresponds to positive values of $a$.
The meson self energy is related to the complex meson-nucleus potential $U(r) = V(r) + i \cdot W(r)$ via 
\begin{eqnarray}
V(r) &=& \frac{{\rm Re} \ \Pi(E,{\vec k},\rho(r))}{2 \cdot E} 
\nonumber \\
W(r) &=& \frac{{\rm Im} \ \Pi(E,{\vec k},\rho(r)) }{2\cdot E}  
\label{eq:eqa3}
\end{eqnarray}
where
$r$ is the distance from the centre of the nucleus.

With a strong attractive interaction there is a chance
to form meson bound states in nuclei \cite{Haider:1986sa}.
If found, these mesic nuclei would be a new state of 
matter bound just by the strong interaction,
without electromagnetic Coulomb 
effects which are absent for the neutral $\eta$ and $\eta'$.
Mesic nuclei differ from mesonic atoms~\cite{Yamazaki:1996zb}
where, for example, a
$\pi^-$ is trapped in the Coulomb potential of the nucleus
and bound by the electromagnetic interaction~\cite{Toki:1989wq}.

For clean observation of a bound state one needs 
larger attraction than absorption and thus the real 
part of the meson-nucleus optical potential to be much 
bigger than the imaginary part.
Does the real part 
provide a sufficiently deep potential? 
Is the imaginary part 
small enough to allow for narrow states that can more easily be detected experimentally?

Studies involving bound state searches and excitation
functions of mesons in photoproduction from nuclear 
targets give information about the $\eta$ and $\eta'$ meson nucleus
optical potentials
\cite{Metag:2017yuh,Bass:2018xmz}.

Strong attractive interactions between the $\eta$ and $\eta'$ mesons and
nucleons mean that both the $\eta$ and $\eta'$ 
are prime targets for mesic nuclei searches,
with a vigorous program of experiments~\cite{Metag:2017yuh}
in Germany and Japan,
plus
equally vigorous theoretical activity.

For the $\eta$, hints for a possible bound state come from a sharp rise in the production cross-sections close to threshold in photoproduction experiments from $^3$He at Mainz 
\cite{Pfeiffer:2003zd,Pheron:2012aj}
and in proton-deuteron, pd, amd deuteron-deuteron, dd,
reactions at COSY 
\cite{Goslawski:2012zz,Budzanowski:2008qx}.
The most precise direct searches (so far) come from the WASA@COSY 
experiment
with focus on possible $^3$He 
\cite{Adlarson:2019haw,Adlarson:2020ldu}
and $^4$He $\eta$ 
\cite{Adlarson:2016dme}
bound states. 
While no clear signal 
is seen within the systematic errors of the experiments, an $^3$He-$\eta$ bound state is not excluded
and tight constraints 
obtained on possible bound state production cross-sections.
Eta bound states in helium require a large $\eta-$nucleon 
scattering length 
with real part greater than about 
0.7--1.1~fm  
\cite{Barnea:2017epo,Barnea:2017oyk,Fix:2017ani}.
New studies of the $\gamma$d reaction
at the ELPH laboratory in Japan
see an interesting structure
in the M$_{\eta d}$ invariant mass distribution close to the $\eta d $ threshold
which might be evidence for an $\eta$ two-nucleon bound state or an $\eta d $ virtual  state due to strong $\eta d$ attraction~\cite{Ishikawa:2021yyz}.

Recent measurements of $\eta'$ photoproduction from 
carbon and niobium nuclear targets have been interpreted 
to imply an effective mass shift $\approx -40$MeV as well as
small $\eta'$ width 
$\approx 13$ MeV in nuclei 
at nuclear matter density 
\cite{Nanova:2012vw,Nanova:2013fxl,Nanova:2016cyn,Nanova:2018zlz,Friedrich:2016cms}
that might give rise to relatively narrow bound 
$\eta'$-nucleus states accessible to experiments.
New experimental groups are looking for possible $\eta'$
bound states in carbon using the (p, d) reaction
at
GSI/FAIR \cite{Tanaka:2016bcp,Tanaka:2017cme,Itahashi},
and photoproduction studies at 
Spring-8 with carbon and copper targets~\cite{Shimizu:2017kua,Tomida:2020yin}.

The plan of this paper is as follows.
Section 2 
highlights the special role of the $\eta$ and $\eta'$ mesons in low energy QCD.
Section 3 
then discusses 
the theory of medium modifications, focusing in Section 3.1 on theoretical predictions for the $\eta$ and $\eta'$ properties in medium.
Next, Section 4 turns to experiments on the $\eta$ in medium with emphasis on new WASA@COSY results and ongoing experiments in Japan.
Section 5 discusses 
Bonn, GSI and Spring-8
measurements of the $\eta'$ in medium and the search for bound states with outlook to new planned experiments at GSI/FAIR.
Finally, Section 6 
summarizes 
the paper 
with an outlook to future experiments.
Complementary reviews of $\eta$ and $\eta'$ interactions with nucleons and nuclei are given in \cite{Metag:2017yuh,Bass:2018xmz}.

\section{\textit{2. The $\eta'$ and $\eta$ mesons with coupling to anomalous glue}}

Spontaneous chiral symmetry breaking means that the chiral symmetry of 
the QCD Lagrangian is broken in the vacuum. 
One finds a non-vanishing chiral
condensate connecting left- and right-handed quarks
\begin{equation}
\langle {\rm vac}  | {\bar \psi} \psi  |  {\rm vac} \ \rangle < 0
.
\label{eqb1}
\end{equation}
This spontaneous symmetry breaking induces an octet of 
light-mass 
pseudoscalar Goldstone bosons associated with SU(3): 
the 
pions and kaons
and also 
(before extra gluonic effects in the singlet channel)
an iso-singlet Goldstone state.

The Goldstone bosons $P$ couple to the axial-vector currents 
which play the role of Noether currents 
through
\begin{equation}
\langle {\rm vac} | J_{\mu 5}^i | P(p) \rangle = 
-i f_P^i \ p_{\mu} e^{-ip.x}
\label{eqb2}
\end{equation}
with $f_P^i$ the corresponding decay constants
(which determine the strength for, 
{\it e.g.}, $\pi^- \to \mu^- {\bar \nu}_{\mu}$)
and satisfy the Gell-Mann-Oakes-Renner (GMOR) relation
\cite{GellMann:1968rz}
\begin{equation}
m_P^2
f_{\pi}^2 = - m_q \langle {\rm vac} | {\bar \psi} \psi | {\rm vac} \rangle 
+ {\cal O} (m_q^2)
\label{eqb3}
\end{equation}
with $f_{\pi} = \sqrt{2} F_{\pi} = 131$ MeV.
The mass squared of the Goldstone bosons $m_P^2$
is in first order proportional to the mass of their 
valence quarks $m_q$.

This picture is 
the starting point of successful pion and kaon phenomenology.
The QCD Hamiltonian is linear in the quark masses.
For small quark masses 
this allows one to perform a rigorous expansion
perturbing in $m_q \propto m_{\pi}^2$,
called the chiral expansion~\cite{Gasser:1982ap}.
The lightest up and down quark masses are determined from detailed studies of chiral dynamics.
One finds
$m_u=2.2^{+0.5}_{-0.3}$ MeV 
and $m_d= 4.7^{+0.5}_{-0.2}$ MeV 
whereas the strange quark mass is 
slightly heavier at $m_s= 93^{+11}_{-5}$ MeV
(with all values here quoted at the scale $\mu = 2$ GeV
 according 
 to the Particle Data Group~\cite{Zyla:2020zbs}).

Whereas Eq.~(\ref{eqb3}) works
very well for the flavour non-singlet pions and kaons, the isosinglet
$\eta$ and $\eta'$ are
more subtle due to gluonic effects in the flavour singlet channel.
The quark condensate in Eq.~(\ref{eqb3}) 
also spontaneously breaks axial 
U(1) symmetry meaning that one might also expect a 
flavour-singlet Goldstone state which mixes with
the octet state to generate the isosinglet bosons.
However, without extra input, the resultant bosons 
do not correspond to states in the physical spectrum.
The lightest mass isosinglet bosons, the $\eta$ and 
$\eta'$, 
with masses $m_{\eta} = 548$ MeV and $m_{\eta'} = 958$ MeV
are about 300-400 MeV too heavy to be pure 
Goldstone states.

The extra ingredient is a gluonic mass term in
the flavour-singlet channel. 
In the singlet channel
the 
quark-antiquark pair (with quark chirality equal two)
propagates with coupling to non-perturbative gluonic intermediate states 
(with zero net chirality);
see Fig.~1.

\begin{figure}[t]
\begin{center}
\setlength{\unitlength}{1mm}
\begin{picture}(36,24)
\put(8,12){\circle*{1}}
\put(16,4){\circle*{1}}
\put(16,20){\circle*{1}}
\put(28,4){\circle*{1}}
\put(28,20){\circle*{1}}
\put(36,12){\circle*{1}}
\put(0,11.5){$\eta'$} 
\put(38,11.5){$\eta'$}
\thicklines
\put(16,4){\line(0,1){16}}
\put(28,4){\line(0,1){16}}
\put(8,12){\line(1,1){8}}
\put(8,12){\line(1,-1){8}}
\put(28,4){\line(1,1){8}}
\put(28,20){\line(1,-1){8}}
\newcommand{\glue}%
{\curve(0,0, 0.5,0.6, 1,0)\curve(1,0, 1.5,-0.6, 2,0)
\curve(2,0, 2.5,0.6, 3,0)\curve(3,0, 3.5,-0.6, 4,0)
\curve(4,0, 4.5,0.6, 5,0)\curve(5,0, 5.5,-0.6, 6,0)
\curve(6,0, 6.5,0.6, 7,0)\curve(7,0, 7.5,-0.6, 8,0)}
\put(16,20){\glue}
\put(20,20){\glue}
\put(16,4){\glue}
\put(20,4){\glue}
\end{picture}
\caption{
Gluonic intermediate states contribute to the $\eta'$.
The $\eta'$ mixes a chirality-two quark-antiquark 
contribution and chirality-zero gluonic contribution.
}
\end{center}
\end{figure}
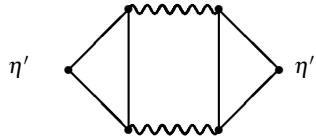

To see the effect of this gluonic mass contribution 
consider the $\eta$-$\eta'$ mass matrix for 
free mesons
with rows and columns in the octet-singlet basis
$
\eta_8 = \frac{1}{\sqrt{6}}\; (u\ubar + d\dbar - 2 s\sbar)$
and
$
\eta_0 = \frac{1}{\sqrt{3}}\; (u\ubar + d\dbar + s\sbar) $.
Expressing these in
terms of the pion and
kaon mass squared,
at
leading order in the chiral expansion
(taking terms proportional to the quark masses $m_q$)
this reads
\begin{equation}
M^2 =
\left(\begin{array}{cc}
{4 \over 3} m_{\rm K}^2 - {1 \over 3} m_{\pi}^2  &
- {2 \over 3} \sqrt{2} (m_{\rm K}^2 - m_{\pi}^2) \\
\\
- {2 \over 3} \sqrt{2} (m_{\rm K}^2 - m_{\pi}^2) &
[ {2 \over 3} m_{\rm K}^2 + {1 \over 3} m_{\pi}^2 + {\tilde m}^2_{\eta_0} ] 
\end{array}\right) .
\label{eqb4}
\end{equation}
Here ${\tilde m}^2_{\eta_0}$ 
is the flavour-singlet gluonic mass term.

The masses of the physical $\eta$ and $\eta'$ mesons are found
by diagonalizing this matrix, {\it viz.}
\begin{eqnarray}
| \eta \rangle &=&
\cos \theta \ | \eta_8 \rangle - \sin \theta \ | \eta_0 \rangle
\nonumber \\
| \eta' \rangle &=&
\sin \theta \ | \eta_8 \rangle + \cos \theta \ | \eta_0 \rangle
\label{eqb5}
\end{eqnarray}
One obtains values for the $\eta$ and $\eta'$ masses:
\begin{equation}
m^2_{\eta', \eta} 
= 
(m_{\rm K}^2 + 
{\tilde m}_{\eta_0}^2 /2)
\pm {1 \over 2}
\sqrt{(2 m_{\rm K}^2 - 2 m_{\pi}^2 - {1 \over 3} {\tilde m}_{\eta_0}^2)^2
   + {8 \over 9} {\tilde m}_{\eta_0}^4} 
.
\label{eqb6}
\end{equation}
Here
the lightest mass state is the $\eta$ and heavier state 
is the $\eta'$.
Summing over the two eigenvalues in Eq.(\ref{eqb6}) gives the 
Witten-Veneziano mass formula
\cite{Witten:1979vv,Veneziano:1979ec}
\begin{equation}
m_{\eta}^2 + m_{\eta'}^2 = 2 m_K^2 + {\tilde m}_{\eta_0}^2 .
\label{eqb7}
\end{equation}
The gluonic mass term is obtained by substituting 
the physical values of 
$m_{\eta}$, $m_{\eta'}$ and $m_K$ 
to give ${\tilde m}_{\eta_0}^2 = 0.73$GeV$^2$. 
In QCD 
${\tilde m}_{\eta_0}^2$
is related 
to a quantity
called the Yang-Mills 
topological susceptibility.
Its value is induced by
non-perturbative gluon dynamics and topological structure in QCD vacuum
associated with the QCD axial anomaly
\cite{Shore:2007yn}
, e.g., instantons and perhaps gluon dynamics related to confinement.

In recent computational QCD lattice calculations
\cite{Cichy:2015jra},
the gluonic term on the right-hand side of Eq.(\ref{eqb7}) 
and 
the meson mass contributions 
(with dynamical quarks)
were both computed.
These  
calculations verified the Witten-Veneziano mass formula
at the 10\% percent level in the QCD lattice approach.

Without the gluonic mass term
the $\eta$ would be approximately an isosinglet light-quark state
(${1 \over \sqrt{2}} | {\bar u} u + {\bar d} d \rangle$)
with mass $m_{\eta} \sim m_{\pi}$
degenerate with the pion and
the $\eta'$ would be a strange-quark state $| {\bar s} s \rangle$
with mass $m_{\eta'} \sim \sqrt{2 m_{\rm K}^2 - m_{\pi}^2}$
--- mirroring the isoscalar vector $\omega$ and $\phi$ mesons.

When interpreted in terms of the leading order mixing scheme,
Eq.~(\ref{eqb5}), phenomenological studies of various decay processes 
give a value for the $\eta$-$\eta'$ mixing angle between 
$-15^\circ$ and $-20^\circ$ \cite{Gilman:1987ax,Ball:1995zv,Ambrosino:2009sc}.
The $\eta'$ has a large flavor-singlet component with 
strong affinity to couple to gluonic degrees of freedom.

In the octet channel the leading order mass term in Eq.~(\ref{eqb4})
before mixing with the
singlet state is the
Gell-Mann Okudo mass term
$
m_{\eta_8}^2 
= \frac{1}{3}
(4 m_K^2 - m_{\pi}^2)
$.
Numerically $m_{\eta_8}$ agrees with
the $\eta$ meson mass 
to within 4\%.
However,
large mixing through the strange quark mass means that non-perturbative glue through axial 
U(1) dynamics plays an important role with both 
 the $\eta$ and $\eta'$ and their interactions. 
 The role of singlet degrees of freedom in the $\eta$ may be essential to understanding the $\eta$-nucleon scattering length $a_{\eta N}$, 
 see Section 3.1.

Besides the meson masses, 
gluonic degrees of freedom are important in axial U(1) dynamics and their effect can be included in an extended effective chiral Lagrangian for low-energy QCD 
\cite{DiVecchia:1980yfw,Witten:1980sp}.
Applications include gluonic
contributions to the $\eta'$-nucleon
coupling constant \cite{Bass:1999is},
the proton's flavour singlet axial-charge 
\cite{Shore:1991dv}
which is related to quark spin content of the proton
\cite{Bass:2004xa,Aidala:2012mv}, 
resonant behaviour in $\eta' \pi$
re-scattering which yields a possible
interpretation of the lightest mass
$1^{-+}$ exotic state 
found with mass in the
range 1400-1600 MeV \cite{Bass:2001zs},
$\eta \to 3 \pi$ decays \cite{Leutwyler:2013wna},
...

So far the $\eta$ and $\eta'$ have been discussed at leading 
order in the chiral expansion.
Going beyond leading order,
one becomes sensitive to extra SU(3) breaking 
through the difference in the pion and kaon decay constants,
$F_{K} = 1.22 F_{\pi}$,
as well as 
new gluonic mediated couplings.
One finds strong mixing also in the decay constants.
Two mixing angles enter the $\eta - \eta'$ system 
when one extends the theory to ${\cal O}(p^4)$ in the meson 
momentum~\cite{Leutwyler:1997yr,Feldmann:1998sh}.

There are several places that glue enters 
$\eta'$ and $\eta$ meson physics:
the gluon topology potential which generates the large 
$\eta'$ and $\eta$ masses, 
possible small mixing with a lightest mass pseudoscalar 
glueball state 
(which comes with a kinetic energy term in its Lagrangian)
and, in high momentum transfer processes, radiatively 
generated glue associated with perturbative QCD.
Possible candidates for the 
pseudoscalar 
glueball state are predicted
by lattice QCD calculations with a mass above 2 GeV 
\cite{Morningstar:1999rf,Gregory:2012hu}.
These different gluonic contributions are distinct physics.

\section{\textit{3. Medium modifications}}

Hadron properties change in medium.
As mentioned in Section 1,
the pion decay constant which acts as an order parameter for
chiral symmetry is suppressed in medium.
Hadron masses and widths are also density (and temperature) dependent.
The study of the QCD phase diagram is one of the main topics in QCD research
\cite{McLerran:2020rnw,
BraunMunzinger:2009zz}
with implications for neutron star structure and the QCD phase transition in the (very hot) early Universe.

This article focuses 
on hadrons at finite nuclear density and zero temperature. 
In medium, 
key issues are
the effect of 
the scalar $\sigma$ (correlated two-pion exchange) mean field as well as $\rho$ and $\omega$
mean fields in the nucleus \cite{Saito:2005rv},
as
well as explicit
pion cloud and rescattering
effects in the nuclear medium~\cite{Ericson:1988gk}.

Medium modifications are observed from low energy
properties 
\cite{Metag:2017yuh}
through to deep inelastic scattering from nuclear targets, which reveals that the quark momentum distributions in the proton are modified when the proton is in a nuclear environment
\cite{Aubert:1983xm,Cloet:2019mql}.

In Gamow-Teller transitions
the proton's isovector axial-charge $g_A^{(3)}$
is observed to be quenched by about 20\% in large nuclei \cite{Ericson:1988gk,
Suhonen:2018ykq}.
Theoretically, this follows from pion cloud effects in the nucleus as well as Ericson-Ericson-Lorentz-Lorenz rescattering 
corrections \cite{Ericson:1998hq,
Ericson:1973vj}.
Through the Bjorken sum-rule \cite{Bjorken:1966jh,Bjorken:1969mm}, 
this quenching of $g_A^{(3)}$ also means that the protons' internal spin structure 
probed in polarized deep inelastic scattering
is expected to be modified in medium \cite{Bass:2020bkl}, 
a result 
found also in partonic models
\cite{deBarbaro:1984gh,Guzey:1999rq,Sobczyk:2000rf,Cloet:2005rt}.
Medium dependence of nucleon spin structure
should persist also to
polarized photoproduction
on polarized nucleons in nuclear targets.
The Gerasimov-Drell-Hearn sum-rule
\cite{Gerasimov:1965et,Drell:1966jv}
relates
the difference 
in the two spin cross-sections
to the ratio of anomalous magnetic moment and nucleon's mass all squared, with both of these terms expected to be medium dependent~\cite{Bass:2020bkl}.

In the medium, besides changes in pion and kaon masses, the anti-proton
effective mass is observed to be reduced by about
100-150 MeV
below their mass in free space
at 2 times nuclear matter density
\cite{Schroter:1994ck}.
Reduction of the nucleon and ${\rm \Delta}$ masses by about -30 MeV have been discussed in 
\cite{Oset:1987re}.
In recent measurements,
the ${\rm \Delta}$ effective mass was observed to be shifted by about -60 MeV
in peripheral and central heavy-ion collisions
\cite{Lenske:2018bgr,Lenske:2018bvq}.
In contrast,
the effective mass of the
N$^*$(1535)
nucleon resonance which couples strongly to the $\eta$ meson 
is observed to be approximately density independent in heavy-ion
collisions \cite{Averbeck:1997ma}
and photoproduction
experiments
\cite{RoebigLandau:1996xa,Yorita:2000bu},
though some evidence for broadening was observed \cite{Yorita:2000bu}.

What should one expect for the $\eta$ and $\eta'$ in medium?
As explain{\bf ed} in Section 3.1,
their masses are expected to be medium dependent with the chance for bound states in light nuclei.

In addition to finite density,
more generally in the QCD phase diagram
axial U(1) symmetry is
also expected to be (partially) restored at finite 
temperature \cite{Kapusta:1995ww}.
This finite temperature result is observed 
in recent QCD lattice calculations 
\cite{Bazavov:2012qja,Tomiya:2016jwr,Aoki:2020noz}.

\subsection{3.1 Modelling the $\eta'$ and $\eta$ in medium}

The $\eta$ and $\eta'$ in medium have been addressed in the context of the mean field
Quark Meson Coupling model,
QMC~\cite{Bass:2005hn,Bass:2013nya}, 
as well as 
chiral coupled channels
\cite{Nagahiro:2011fi},
Nambu-Jona-Lasinio 
~\cite{Nagahiro:2006dr,Bernard:1987sx},
and linear $\sigma$ model
\cite{Sakai:2013nba}
calculations.

This section focuses on the QMC approach,
which
predicts an
$\eta'$ effective mass shift of $\approx$ -37 MeV at nuclear matter density $\rho_0$,  
the one model prediction
very similar to the 
results of the CBELSA/TAPS experiment discussed in Section 5
below~\cite{Bass:2005hn,Bass:2013nya}.
In the QMC model medium modifications are calculated at
the quark level through coupling of the light quarks in 
the hadron to the scalar isoscalar $\sigma$ 
(and also $\omega$ and $\rho$) mean fields in the nucleus
\cite{Saito:2005rv,Guichon:1987jp,
Guichon:1995ue}.
One works in mean field approximation.
The couplings of light-quarks 
to the $\sigma$ (and $\omega$ and $\rho$) mean fields in 
the nucleus are adjusted to fit the saturation energy and 
density of symmetric nuclear matter and the bulk symmetry energy.

The large $\eta$ and $\eta'$ masses are used to motivate 
taking a MIT Bag description for the meson wavefunctions,
\cite{Tsushima:1998qw,Tsushima:1998qp}.
Phenomenologically, the MIT Bag gives a good fit
 to meson properties in free space for the kaons 
 and heavier hadrons \cite{DeGrand:1975cf}.
Gluonic topological effects are understood to be 
``frozen in'', meaning that they are only present 
implicitly through the masses and mixing angle in the model.
The strange-quark component of the wavefunction does not 
couple to the $\sigma$ mean field and $\eta$-$\eta'$ mixing 
is readily built into the model.
Possible binding energies and the in-medium masses of 
the $\eta$ and $\eta'$ are sensitive to the flavor-singlet component in the mesons 
and hence to the non-perturbative 
glue associated with axial U(1) dynamics~\cite{Bass:2005hn}.
Working with the mixing scheme in Eq.~(\ref{eqb5})
with an $\eta$-$\eta'$ mixing angle of $-20^\circ$ 
the QMC prediction for the $\eta'$ mass in medium at nuclear 
matter density is 921 MeV, that is a mass shift of $-37$ MeV. 
This value is in excellent agreement with the mass shift 
$-40 \pm 6 \pm 15$~MeV deduced from photoproduction data,
see
Eq.~(\ref{eqe4})
in Section 5. 
Mixing increases the octet relative to singlet component in
the $\eta'$, reducing the binding through increased strange
quark component in the $\eta'$ wavefunction.
Without the gluonic mass contribution the $\eta'$ 
would be a strange quark state after $\eta$-$\eta'$ mixing.
Within the QMC model there would be no coupling to the 
$\sigma$ mean field and no mass shift so that any observed mass shift 
is induced by 
non-perturbative
glue 
that generates part of the $\eta'$ mass.

For the $\eta$ meson the potential depth predicted by QMC 
is $\approx -100$~MeV at nuclear matter density with 
-20 degrees mixing. 
For a pure octet $\eta$ the model predicts a mass shift of
$\approx -50$~MeV.
Increasing the flavor-singlet component in the $\eta$ at 
the expense of the octet component gives more attraction, 
more binding and a larger value of the $\eta$-nucleon scattering length, $a_{\eta N}$.

The mass shifts obtained in the QMC model with mixing angle -20 degrees correspond to meson-nucleon scattering lengths 
with real parts
Re $a_{\eta N} \approx 0.85$ fm and
Re $a_{\eta' N} \approx 0.47$ fm.
These values are quoted in
linear density approximation with the
Ericson-Ericson-Lorentz-Lorenz denominator switched off (corresponding to the model mean field approximation).
The QMC model makes no statement about the imaginary parts of the meson-nucleus potentials and the corresponding scattering lengths.

In QMC $\eta$-$\eta'$ mixing with the phenomenological 
mixing angle $-20^\circ$ leads to a factor of 
two increase in the mass-shift and 
in the scattering length obtained in the model
relative to the prediction for a pure octet $\eta_8$
\cite{Bass:2005hn}.
This result may explain why values of $a_{\eta N}$ 
extracted from phenomenological fits to experimental 
data where the $\eta$-$\eta'$ mixing angle is unconstrained 
give larger values (with real part about 0.9 fm)
than those predicted 
in theoretical coupled-channel models where the $\eta$ 
is treated as a pure octet state.

The QMC model results for
the $\eta$ and $\eta'$ mass shifts with mixing angle
-20 degrees and for
nuclear densities 
$\rho$
between about 0.5 and 1 times $\rho_0$
are
\cite{Bass:2005hn}
\begin{eqnarray}
m_{\eta}^*/m_{\eta} 
&\approx& 
1 - 0.17 \ \rho/\rho_0
\\
\nonumber
m_{\eta'}^*/m_{\eta'} 
&\approx& 
1 - 0.05 \ \rho/\rho_0 .
\end{eqnarray}

More generally, within the U(1) extended effective chiral Lagrangian approach,
one can
couple the gluonic degrees of freedom 
associated with 
the square of the
topological charge density that gives
the large value of
$\tilde{m}^2_{\eta_0}$
to the $\sigma$ mean field in the nucleus.
This yields a reduced value for 
$\tilde{m}^2_{\eta_0}$
independent of the sign of the coupling
\cite{Bass:2005hn}.
In the QMC approach
if one assumes that the mass formula Eq.~(\ref{eqb4}) holds also in symmetric nuclear
matter at finite density and substitute the QMC predictions for the 
$\eta'$, $\eta$ 
kaon masses in medium (
$m^*_K
= 430.4$ MeV), then ome obtains 
$\tilde{m}^2_{\eta_0}=
0.68$ GeV$^2 < 0.73$ GeV$^2$ 
at $\rho_0$
with $\eta - \eta'$
mixing angle equal 
to -20$^\circ$.

Recent coupled-channel model calculations have appeared 
with mixing and vector meson channels included, 
with predictions for $\eta'$ bound states 
for a range of possible values of $a_{\eta' N}$ 
\cite{Nagahiro:2011fi}.
Larger mass shifts, downwards by up to 80-150 MeV, 
were found in Nambu-Jona-Lasinio model calculations 
(without confinement) 
~\cite{Nagahiro:2006dr}
and in linear sigma model calculations 
(in a hadronic basis) 
\cite{Sakai:2013nba}
which also gave a rising $\eta$ effective mass at finite density.
An early calculation
\cite{Bernard:1987sx}
gave close to zero effect for the $\eta'$.

For the $\eta'$-nucleon scattering length,
at tree level the flavour-singlet version of the Weinberg-Tomozawa term has 
$a_{\eta'N}$
proportional to the meson mass squared, which
does not vanish in the chiral limit due to the gluonic contribution 
${\tilde m}^2_{\eta_0}$
to the
$\eta'$
mass squared \cite{Bass:2018xmz,Bass:2010kr}.
To this level,
one finds
a finite value for 
the real part of $a_{\eta'N}$.
This situation contrasts with the isovector scattering length $a_{\pi N}$ where the pion mass squared vanishes with zero quark masses.

For possible $\eta$ bound states in nuclei, there are a host of predictions.
Experiments have so far focused on light helium nuclei, with the status discussed in detail
in Section 4 below.
Predictions for heavier nuclei are given in
Refs.~\cite{Tsushima:1998qw,Tsushima:1998qp,Garcia-Recio:2002xog,Friedman:2013zfa,Cieply:2013sga}.

In the following sections these theoretical considerations are confronted with experimental observations.

\section{\textit{4. The $\eta$ -nucleus interaction and the search for $\eta$ mesic states}}

 While the interaction of pions and kaons  with nucleons and nuclei can be studied experimentally using beams of these mesons, $\eta$ and $\eta'$ mesons with lifetimes $\le 10^{-19}$~s are too short-lived to produce particle beams. Information on the $\eta,\eta'$ interaction with nucleons and nuclei can thus only be deduced from final state interactions in the production of these mesons off nucleons and nuclei. As pointed out in the introduction, the meson-nucleus potential can be determined from measurements of the meson-nucleon scattering length, the mass shift and the absorption of these mesons in the nuclear medium, as well as by the measurements of the energies and widths of the meson-nucleus bound states. 
 
 The interaction of $\eta^{\prime}$ with nucleons and nuclei will be discussed in the next section.  
 Here 
  the focus is on the $\eta$-nucleon interaction that was studied based on the energy dependence of the near-threshold photo- and hadro-production of $\eta$ meson off nucleons and nuclei. 

As explained in the Introduction the $\eta$-nucleon complex optical potential is related to the width and binding energy of the $\eta$-nucleus bound state referred to as $\eta$ mesic nucleus. 
Based on theoretical estimations that the $\eta$-nucleon potential is attractive~\cite{Bahlero}, Haider and Liu~\cite{Haider:1986sa} postulated the existence of $\eta$-mesic nuclei. 
This inspired much experimental work
though early
experimental searches for such states (using photon~\cite{Pheron:2012aj,Baskov:2012yd}, 
pion~\cite{Chrien:1988gn},
proton~\cite{Adlarson:2019haw,Adlarson:2020ldu,Budzanowski:2008fr} 
or deuteron~\cite{Adlarson:2016dme,Afanasiev,Moskal-Smyrski,WASA-pi-} 
beams) were not successful and only upper limits for the production of $\eta$-mesic nuclei, and hence only limits on
possible values of the $\eta$-nucleon potential parameters
have been determined.

Among these early experiments, 
the COSY-GEM Collaboration
\cite{Budzanowski:2008fr}
studied the reaction
${\rm p \ ^{27}Al \to ^3 He \ p \pi^- X}$
at recoil free kinematics.
If an $\eta$ meson were produced here 
it would be almost at rest in the laboratory system with chance to be bound. 
The disintegration of the $\eta$ mesic state may then occur via the excitation of a N*(1535) resonance decaying into a back-to-back ${\rm \pi^- p}$ pair.
Some enhancement was observed, 
though this could also be associated with an excited $^{25}$Mg intermediate state. 
An upper bound for the bound state production cross section of about 0.5 nb was deduced.

\begin{figure}[h!]
        \centerline{
        \includegraphics[width=6.8cm,height=5cm]{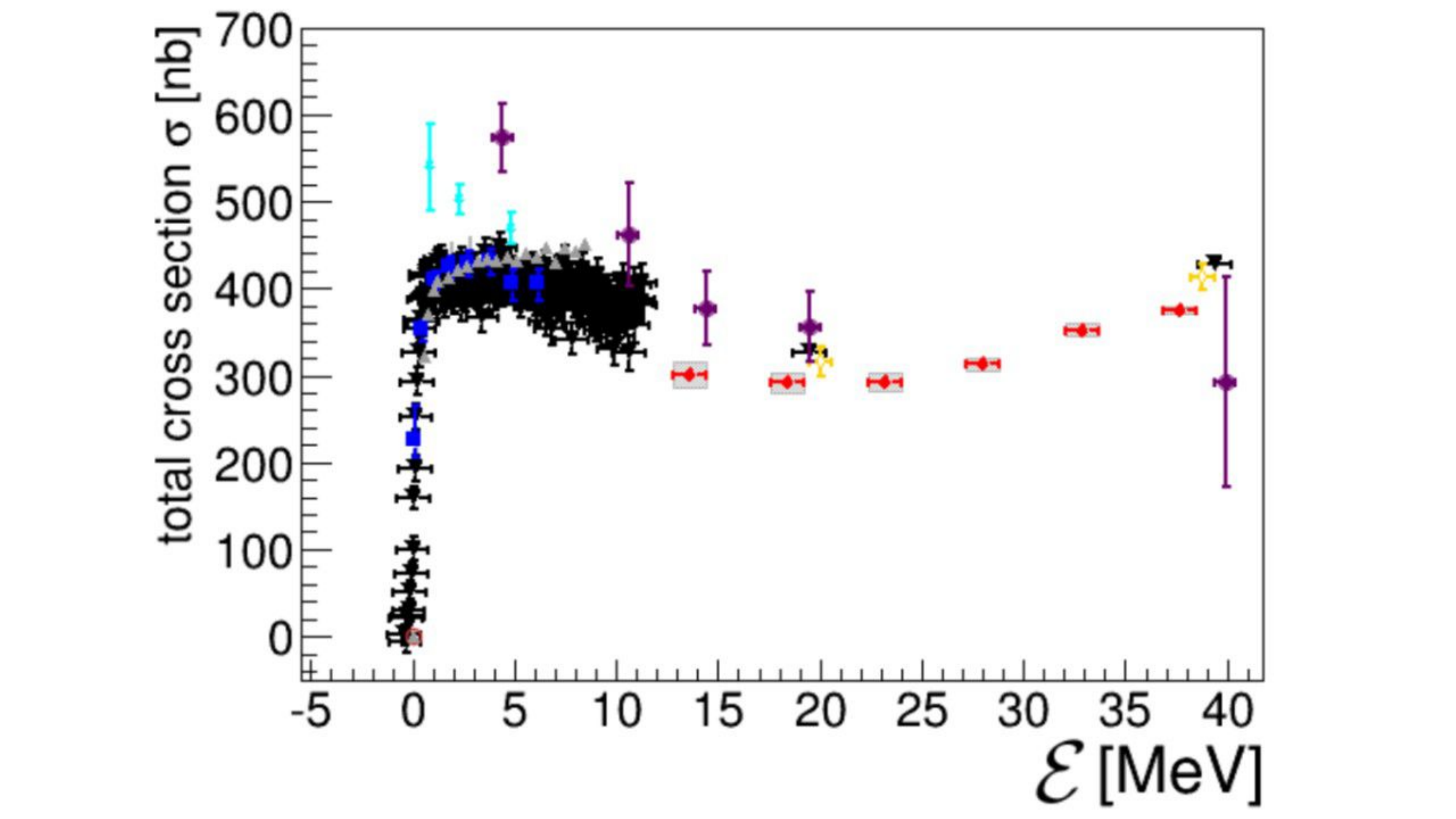}
        \includegraphics[width=5.8cm,height=5cm]{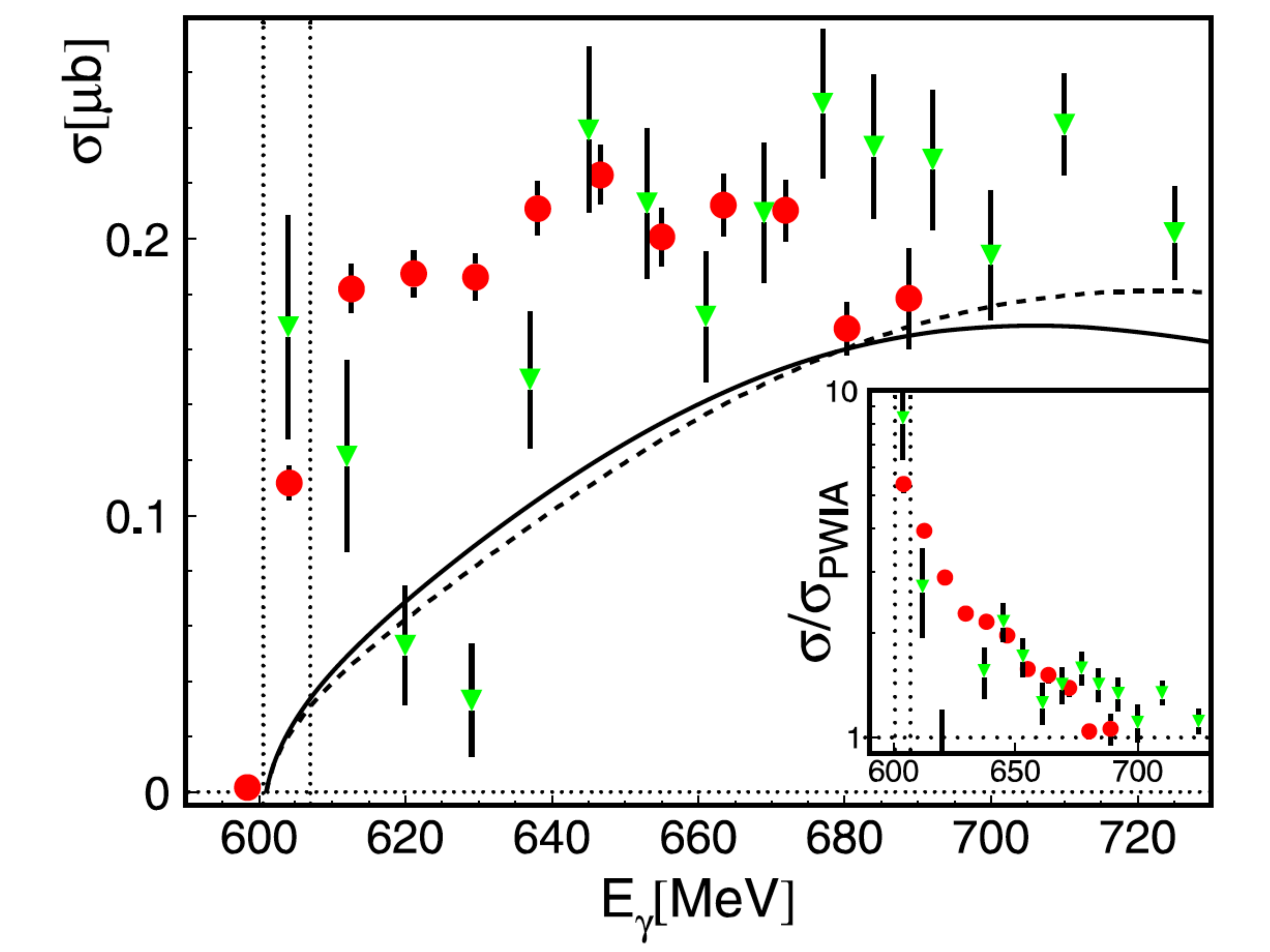}}
\caption{ 
(Left) Total cross section for the ${\rm pd\rightarrow {}^3\text{He}\,\eta}$
reaction as a function of the excitation energy. Symbols represents experimental results described in Refs.~\cite{Berger:1988ba,Mayer:1995nu,Betigeri:1999qa,Bilger:2002aw,Smyrski:2007nu,Adam:2007gz,Mersmann:2007gw,Rausmann:2009dn,Adlarson:2014ysb,Adlarson:2018rgs}. 
The sharp rise of the cross section is visible in the short range of about 2~MeV above the reaction threshold. 
The figure is taken from~\cite{Bass:2018xmz}.
(Right) 
Total cross section for the $\gamma \ ^3$He$ \rightarrow \eta ^3$He reaction shown as a function of the excess energy~\cite{Pheron:2012aj,Pfeiffer:2003zd}.
Solid (dashed) 
Superimposed curves indicate results obtained based on the plane wave impulse approximation (PWIA) 
calculations with a realistic (solid) and isotropic (dashed) angular distribution for the ${\rm \gamma n \rightarrow n \eta}$ reaction. 
Insert: ratio of measured and PWIA cross sections. 
The figure is taken from 
\cite{Pheron:2012aj}.
\label{fig:3He-photo-totalcross}
}
\end{figure}

The most resent high statistics experiments performed by the WASA-at-COSY collaboration~\cite{WASA-pi-,Adlarson:2019haw,Adlarson:2020ldu,Adlarson:2016dme} 
aimed at the observation of the $^4$He-$\eta$ and $^3$He-$\eta$ mesic nuclei. These experiments were motivated by 
hints of a strong $^4$He-$\eta$ and even stronger $^3$He-$\eta$ interaction indicated by the steep growth of the cross sections at the thresholds for the ${\rm dd\to^4\!\!He\eta}$~\cite{Budzanowski:2008qx,Wronska,Willis,Frascaria}, 
${\rm \gamma \ ^3He~\to
^3\!\!He \ \eta}$  
and 
${\rm pd\to ^3\!\!He\eta}$ reactions. Fig.\ref{fig:3He-photo-totalcross} presents cross section excitation functions for the case of 
$^3$He-$\eta$. The steep rise of the cross sections observed for both photo- and hadro-production reactions indicates that the effect is due to the He-$\eta$ final state interaction rather than the initial state reaction dynamics. This conclusion was strengthened by a small and energy independent value of the analyzing power~\cite{Papenbrock:2014hup}
as well as strong variation with energy of the
phase of the 
$s$-wave production amplitude 
~\cite{Smyrski:2007nu,Mersmann:2007gw,Wilkin:2007aa}
as expected for a bound or virtual ${^{3}\mbox{He}}-\eta$ state~\cite{Wilkin:2007aa}.
Based on the above discussed experimental hints,
the WASA-at-COSY experimental searches have focused on possible $\eta$ bound states in $^3$He and $^4$He~\cite{Adlarson:2019haw,Adlarson:2020ldu,Adlarson:2016dme,WASA-pi-}. Also theoretically the existence of the $\eta$ bound states in helium is predicted, provided that the real part of the $\eta-$nucleon 
scattering length is greater than about 
0.7--1.1~fm~\cite{Barnea:2017epo,Barnea:2017oyk,Fix:2017ani}. This requirement overlaps with the range of values from  0.18~fm~\cite{Liu2007} 
up to 1.07~fm~\cite{Green1997} 
predicted for the real part of the $\eta$-nucleon scattering length. The relatively large range of predicted values is due to the different analysis methods.
The smallest values, in the order of 0.2~fm result from 
chiral coupled channel models where the $\eta$ meson is treated in pure octet approximation. Moreover, for most of 
the coupled channels
analyses the imaginary part is larger then the real one (e.g.~$a_{\eta {\rm N}} =$0.18+i0.42~fm~\cite{Liu2007}),
which would imply
that the $\eta$-mesic nucleus cannot exist. 
On the other hand analysis of the experimental data ($\pi {\rm N} \rightarrow \pi {\rm N}$, 
$\pi {\rm N} \rightarrow \eta {\rm N}$, $\gamma {\rm N} \rightarrow \pi {\rm N}$, $\gamma {\rm N} \rightarrow \eta {\rm N}$) in the frame  
of coupled $\eta {\rm N}$,  $\pi {\rm N}$,  $\gamma {\rm N}$ systems,  described by a K-matrix, result in scattering lengths even as large as  
$a_{\eta {\rm N}}=$1.07~fm~+~i0.26~fm~with imaginary part much less than the real one~\cite{Green1997} indicating favorable conditions for the creation of the $\eta$-mesic nucleus. 
Compilations of values derived for the $\eta$-nucleon scattering length in different approaches 
may be found in references~\cite{Sibirtsev2002,Arndt:2005dg}.

\begin{figure}[h!]
\centering
\includegraphics[width=0.85\textwidth]{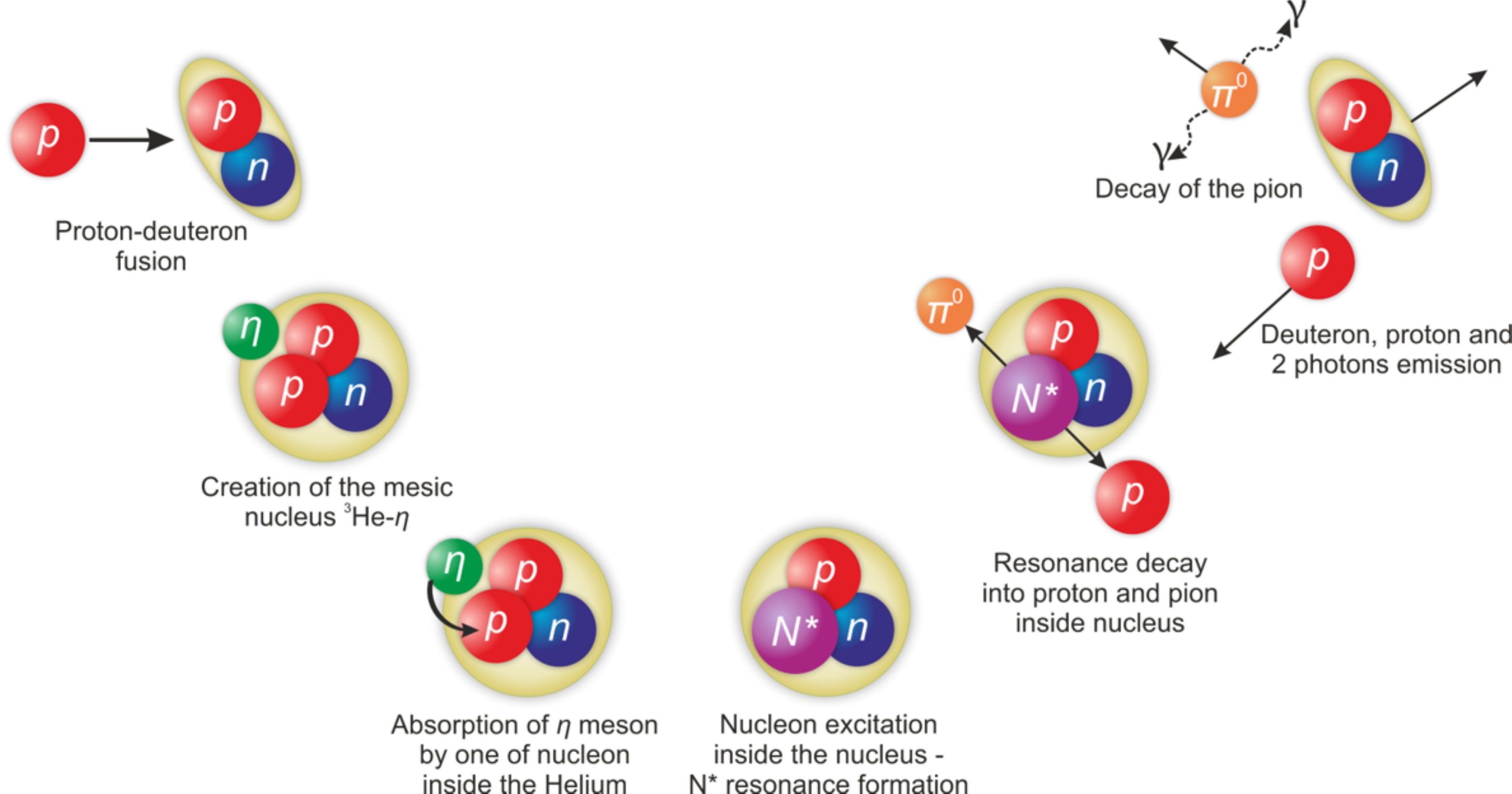}
\caption{ Sequence of processes leading to the 
 $^{3}\hspace{-0.03cm}\textrm{He-}\eta$ bound state production and decay in the ${\rm pd} \rightarrow {\rm d p} \pi^{0}$ reaction.
 The figure is taken from Ref.~\cite{Adlarson:2020ldu}.
~\label{reaction-scheme}
}
\end{figure}

\begin{figure}[h!]
\centering
\includegraphics[width=5.8cm,height=5.0cm]{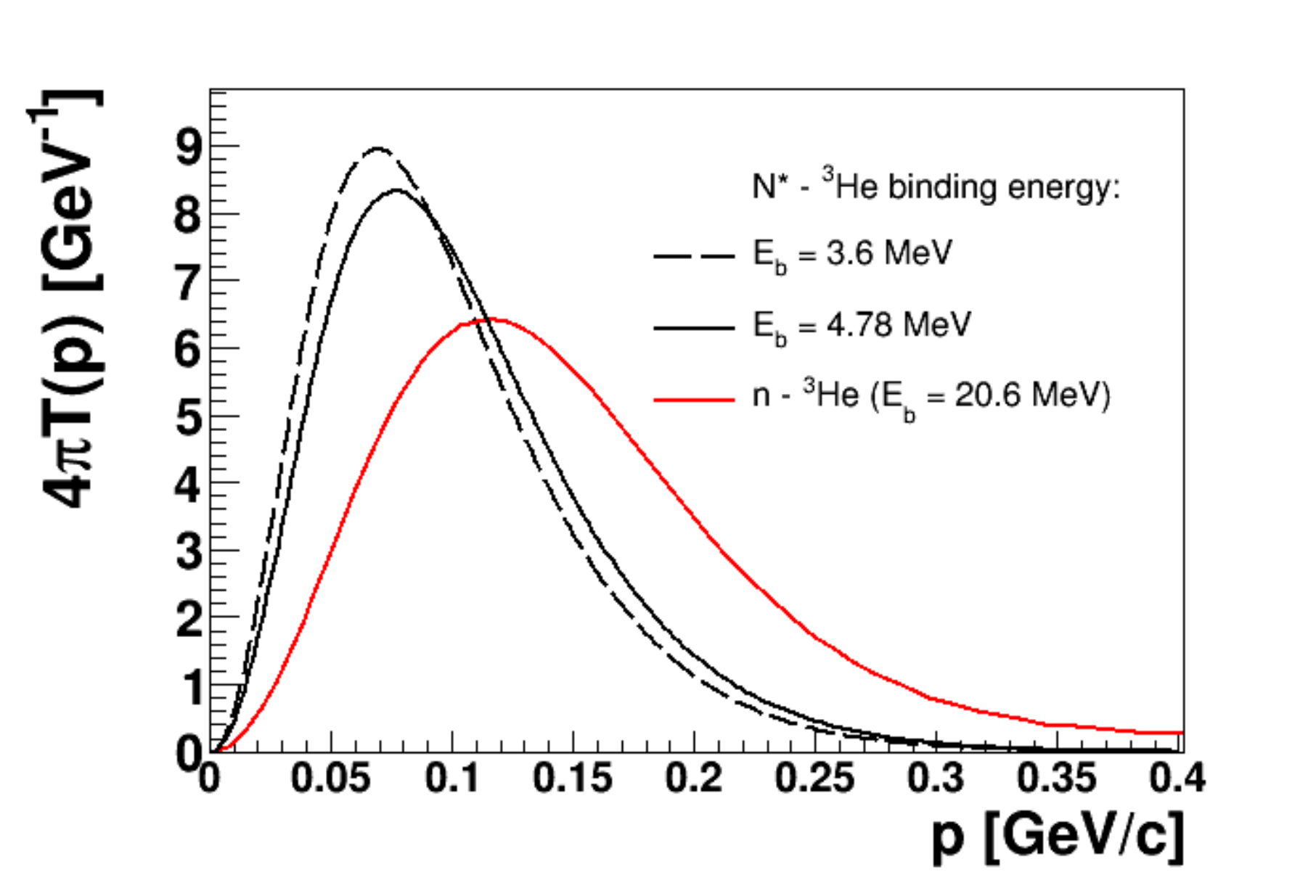}
\includegraphics[width=5.8cm,height=5.0cm]{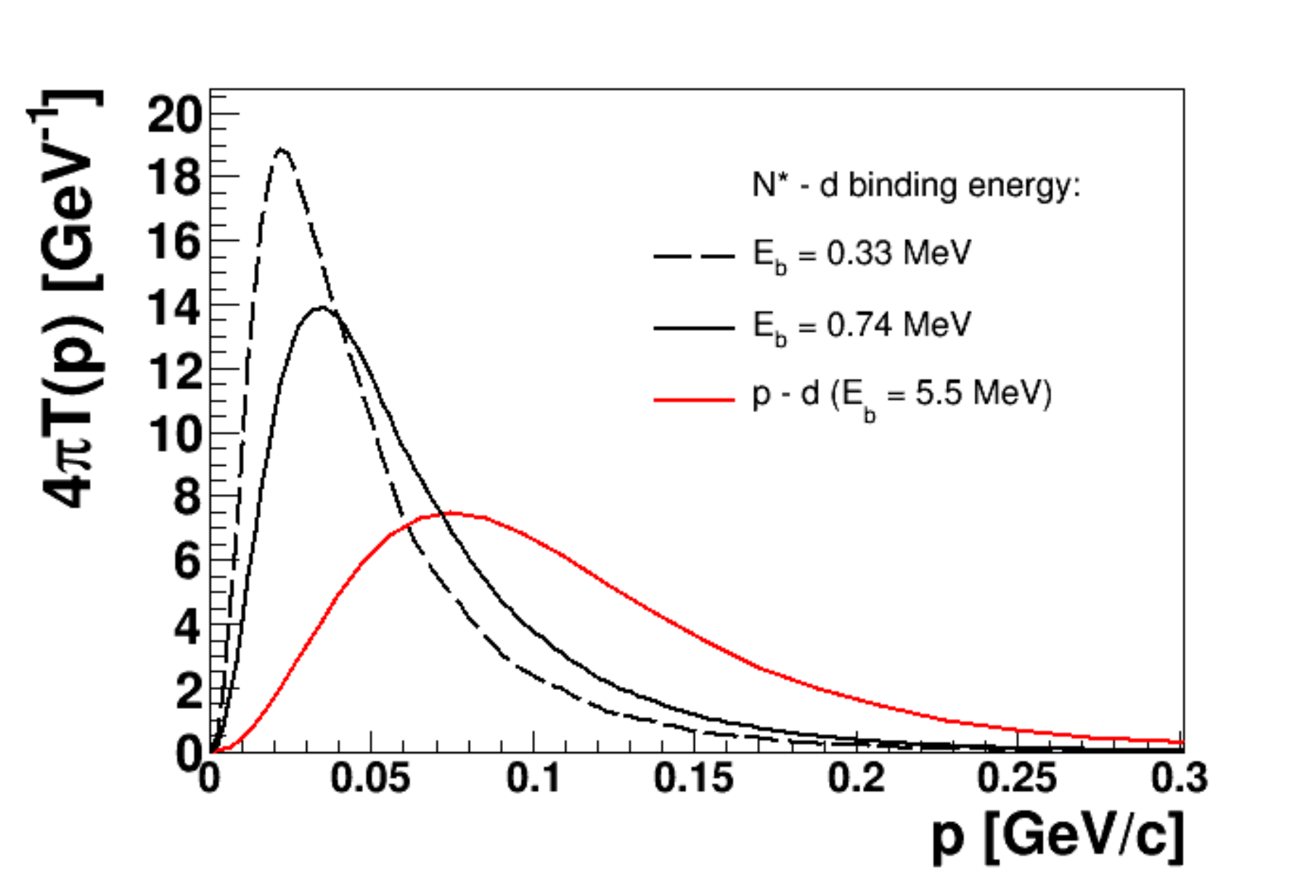}
\caption{ (Left) Momentum distribution of nucleons and N$^{*}$ inside a $^{4}\hspace{-0.03cm}\mbox{He}$ calculated assuming N$^{*}$-$^{3}\hspace{-0.03cm}\mbox{He}$ binding energies of 3.6 MeV and 4.78 MeV, and a $n$-$^{3}\hspace{-0.03cm}\mbox{He}$ separation energy of 20.6 MeV. 
(Right) Momentum distribution of the N$^{*}$ and nucleons inside a $^{3}\hspace{-0.03cm}\mbox{He}$ calculated assuming N$^{*}$-d binding energies of 0.74~MeV and 0.33~MeV, and a \mbox{p-$^{3}\hspace{-0.03cm}\mbox{He}$} potential giving proton separation energy of 5.5 MeV (red solid line). The figures are  obtained based on Refs.~\cite{KelkarEPJA2016,KelkarIJMPE2019,KelkarNPA2020}, and are taken from~\cite{Magda-rev}. With kind permission of Springer.
\label{Nstar_distr}
}  
\end{figure}

\begin{figure}[h!]
\centering
\includegraphics[width=5.8cm,height=4.0cm]{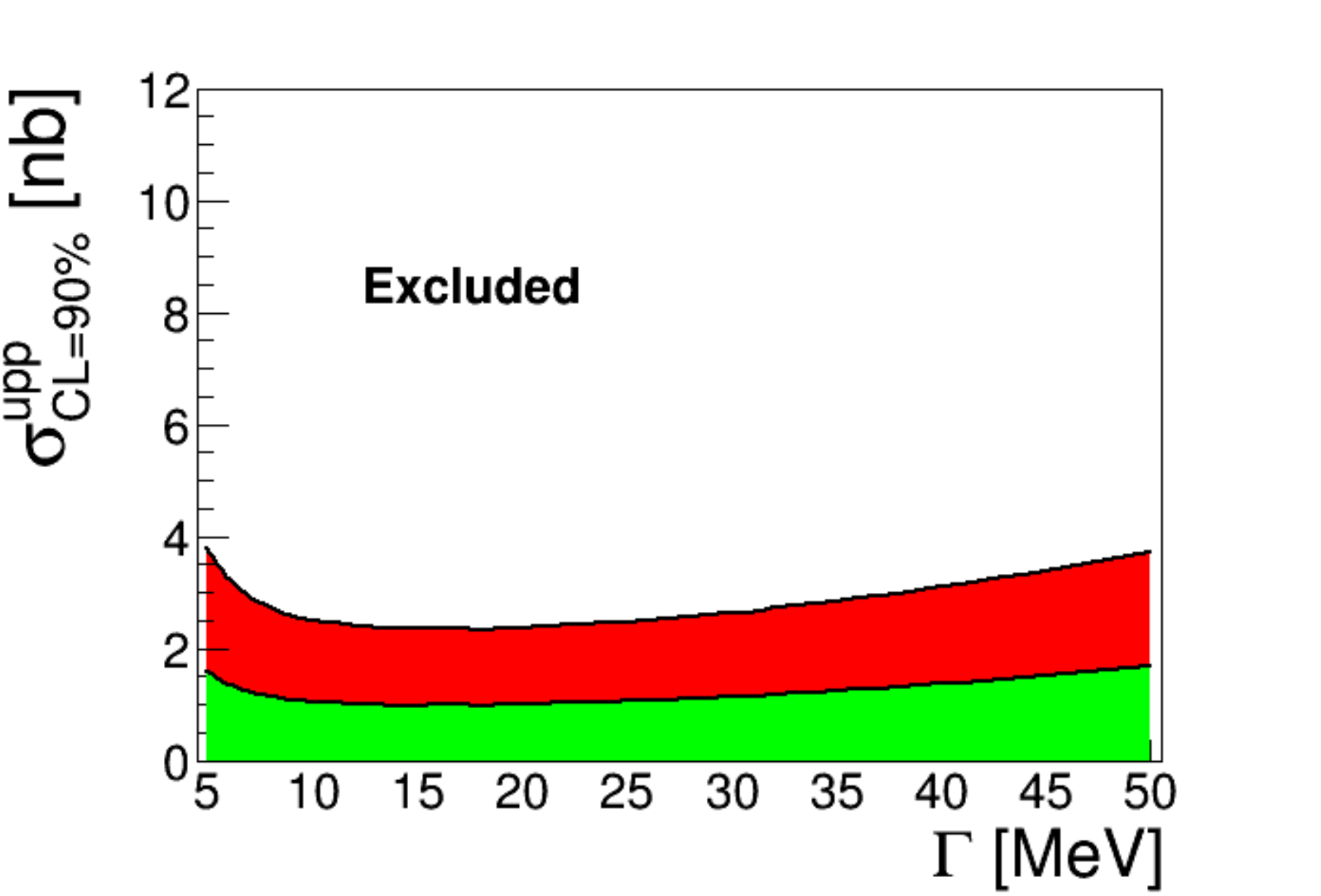}
\includegraphics[width=5.8cm,height=4.0cm]{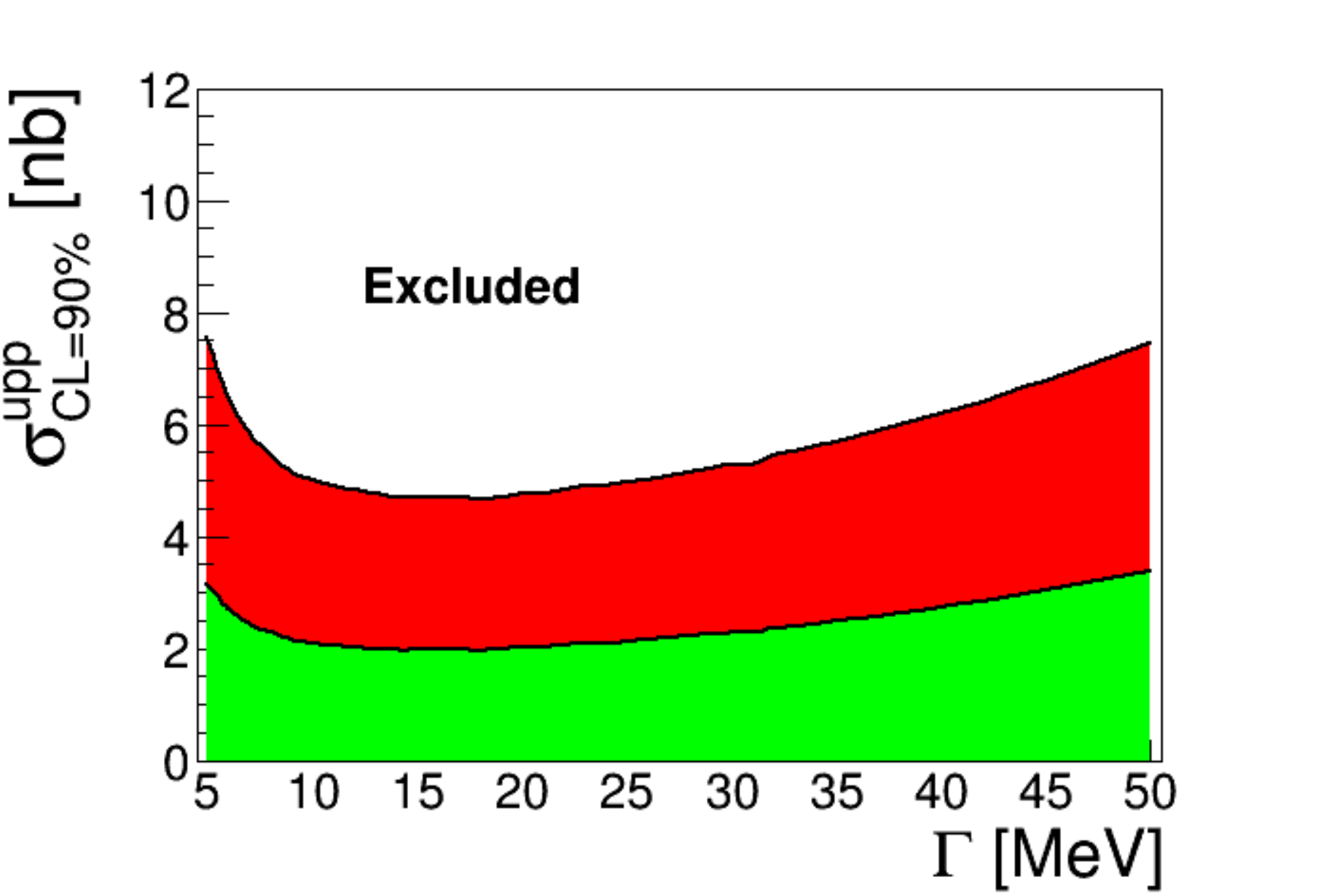}
\caption{
Upper limit of the total cross-section shown as a function of the width of the bound state for the  ${\rm dd} \rightarrow(^{4}\hspace{-0.03cm}\mbox{He}$-$\eta)_{\rm bound}\rightarrow$ $^{3}\hspace{-0.03cm}\mbox{He} n \pi{}^{0}$ (left panel) and the ${\rm dd} \rightarrow(^{4}\hspace{-0.03cm}\mbox{He}$-$\eta)_{\rm bound}\rightarrow$ $^{3}\hspace{-0.03cm}\mbox{He} p \pi{}^{-}$ (right panel). The values were obtained for the binding energy equal to 30~MeV. The result was determined via the simultaneous fit for both channels. The green area denotes the systematic uncertainties. The figures are taken from~\cite{Adlarson:2016dme}.
\label{Result_sigma_upp_both}
}
\end{figure}

\begin{figure}[h!]
\centering
\includegraphics[width=5.8cm,height=5.0cm]{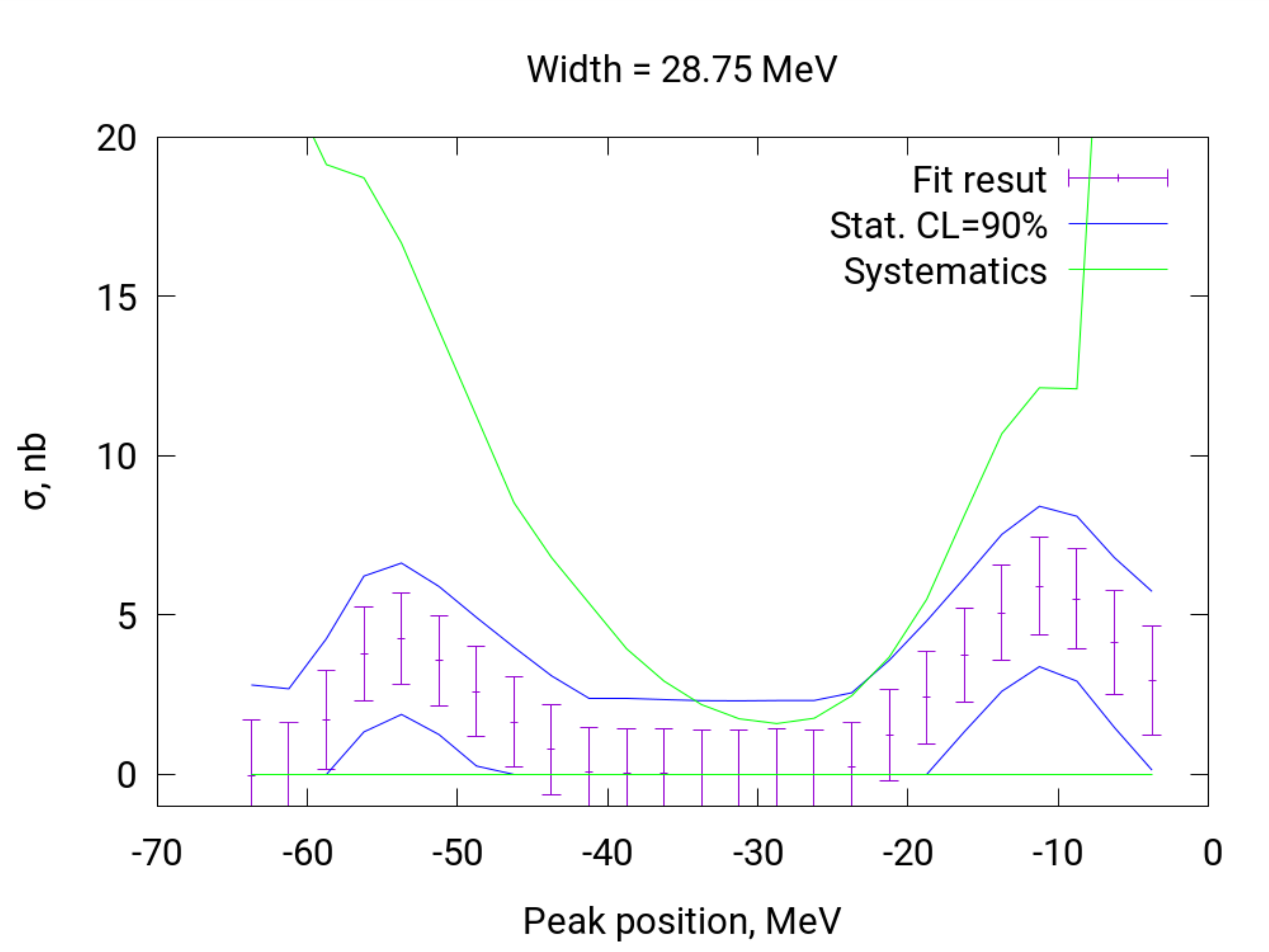}
\includegraphics[width=5.8cm,height=5.0cm]{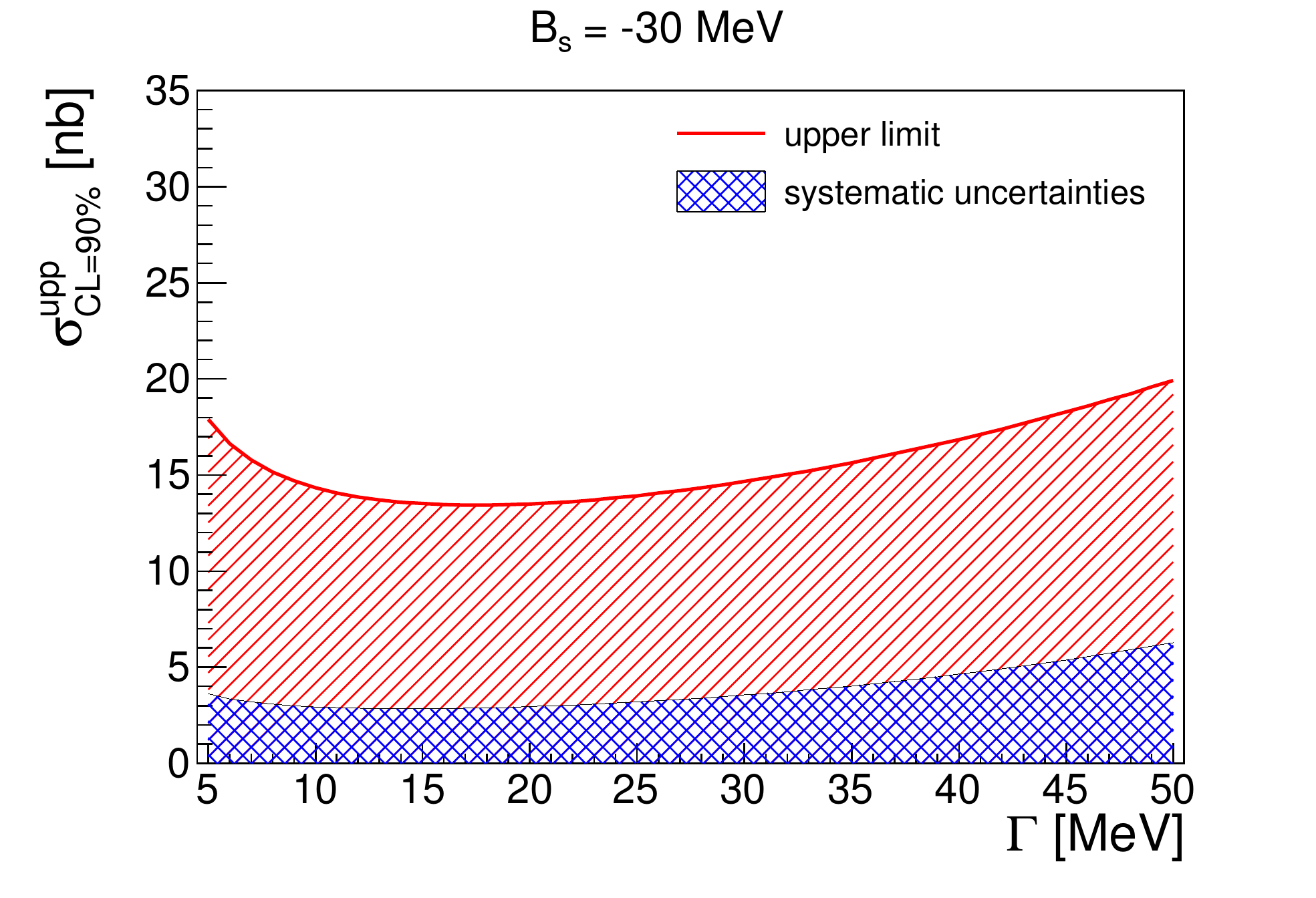}
\caption{
(Left) Upper limits for cross section of the reaction chain ${\rm pd} \rightarrow$ ($^{3}\hspace{-0.03cm}\mbox{He}$-$\eta)_{\rm bound} \rightarrow$ $^{3}\hspace{-0.03cm}\mbox{He}$($\eta~ \mbox{decays})$ as function of the binding energy, assuming in the analysis the width $\rm \Gamma$=28.75~MeV. The blue and green lines show the range of possible bound state production cross section including  statistical and systematic uncertainty respectively. 
The figure is taken from Ref.~\cite{Adlarson:2019haw}.
(Right) 
    The upper limit (90\% CL) of the total cross section for formation of the $^{3}\hspace{-0.03cm}\textrm{He-}\eta$ bound state and its decay via the $pd\rightarrow (^{3}\hspace{-0.03cm}\textrm{He-}\eta)_{\rm bound} \rightarrow {\rm d p} \pi^{0}$ reaction as a function of the width of the bound state. The result for the binding energy of $B_{s} = -30$~MeV is shown. 
	The blue checkered area at the bottom represents the systematic uncertainties.
	The figure is taken from Ref.~\cite{Adlarson:2020ldu}.
\label{upper_limit_1d}
}
\end{figure}

The key physical process of the $\eta$-mesic helium involves a virtual $\eta$ meson production forming a bound state with the helium nucleus in which it is produced. The process is 
illustrated in Fig.~\ref{reaction-scheme}
for the example of the proton-deuteron reaction. The WASA-at-COSY experiment tested two possible decay mechanisms of the $\eta$ mesic helium. In the first scenario (as shown in Fig.~\ref{reaction-scheme}) the $\eta$ meson is absorbed by the nucleon exciting it to the N$^*$(1535) resonance. Subsequently the $N^*$ decays into a pion-nucleon pair leading to the disintegration of the mesic-nucleus. For this decay mechanism three reactions were studied: 
${\rm dd} \to (^4\!{\rm He}-\eta)_{\rm bound}\to ^3\!\!\!{\rm He}\, {\rm p}\, \pi^-$~\cite{WASA-pi-}, 
${\rm dd} \to (^4\!{\rm He}-\eta)_{\rm bound}\to ^3\!\!\!{\rm He}\, {\rm n}\, \pi^0\to ^3\!\!{\rm He}\, {\rm n}\, \gamma\gamma$~\cite{Adlarson:2016dme}, 
and ${\rm pd} \to (^3\!{\rm He}-\eta)_{\rm bound}\to {\rm d p} \pi^0 \to {\rm d}\, {\rm p}\, \gamma\gamma$~\cite{Adlarson:2020ldu}. The latter process is shown in Fig.~\ref{reaction-scheme}. 
In the second considered mechanism the virtual $\eta$ meson is decaying directly, leaving the remaining helium nucleus intact. To test this mechanisms the ${\rm pd}\to (^3\!{\rm He}-\eta)_{\rm bound}\to ^3\!\!{\rm He}\, 3\pi^0 \to ^3\!\!{\rm He}\, 6\gamma$ and ${\rm pd}\to (^3\!{\rm He}-\eta)_{\rm bound}\to ^3\!\!{\rm He}\, 2\gamma$ reaction chains were studied~\cite{Adlarson:2019haw}.
In the case of the decay process proceeding via creation of the N$^*$, due to the finite geometrical acceptance of the WASA-at-COSY detector~\cite{HHAdam},
the determination of the cross section for the studied reactions required the knowledge of the momentum distribution of the N$^*$ resonance inside the mesic-nucleus, and in the case of the second considered mechanism the knowledge about the Fermi momentum distribution for a bound $\eta$ meson orbiting around the $^3$He nucleus is required. The latter was estimated for various combinations
of the $^3$He-$\eta$ optical potential parameters in reference~\cite{Skurzok-NP}.
Fig.~\ref{Nstar_distr} presents the N$^*$ momentum distribution in the 
N$^{*}$-$^3$He and N$^*$NN systems as estimated recently in Refs.~\cite{KelkarEPJA2016,KelkarIJMPE2019,KelkarNPA2020}. These distributions were obtained based on the elementary ${\rm NN}^* \to {\rm NN}^*$ amplitudes within a pion plus $\eta$ meson exchange model. 
In Fig.~\ref{Nstar_distr} the ${\rm N}^*$-$^3\!{\rm He}$ and N$^*$-d momentum distributions are presented for the two chosen binding energies as it is indicated in the legend. The separation energy of nucleons in $^4$He and $^3$He is much larger than the binding energy of N$^*$, therefore the N$^*$ momentum distribution is narrower with respect to the distribution of 
nucleons in helium, as it is visible in Fig.~\ref{Nstar_distr}. 
As the result of the performed experiments and the data analysis, the excitation functions around the $\eta$ meson production thresholds were established for the 
${\rm dd}\to ^3\!\!{\rm He}\, {\rm p}\, \pi^-$~\cite{WASA-pi-}, 
${\rm dd} \to ^3\!\!{\rm He}\, {\rm n}\, \pi^0$~\cite{Adlarson:2016dme}, 
${\rm pd}\to {\rm d}\, {\rm p}\,\pi^0\to {\rm d}\, {\rm p}\, \gamma\gamma$~\cite{Adlarson:2020ldu}, and
${\rm pd} \to ^3\!\!{\rm He}\, 2\gamma$ and ${\rm pd} \to ^3\!\!{\rm He}\, 6\gamma$~\cite{Adlarson:2019haw}.
If the production cross section for the creation of the $\eta$-helium bound state 
 were larger than the achieved experimental uncertainty, then the bound state would manifest itself as a resonance structure on the excitation function, below the $\eta$ meson production threshold. 
Yet, all determined excitation functions were smooth within the error bars and have not revealed any structure which could have been assigned to the formation of 
$\eta$-mesic helium~\cite{Adlarson:2019haw,Adlarson:2020ldu,Adlarson:2016dme,WASA-pi-}.
The upper limits of the cross sections for the production and decay of 
$\eta$-mesic $^4$He and $\eta$-mesic $^3$He are presented in Fig.~\ref{Result_sigma_upp_both} 
and Fig.~\ref{upper_limit_1d}, respectively.

The achieved experimental sensitivity, 
$\sim$6~nb for the ${\rm dd}\to (^4\!{\rm He}-\eta)_{\rm bound}\to \ ^3\!{\rm He}\, {\rm p}\, \pi^-$ process, and
 $\sim$3~nb for the ${\rm dd}\to (^4\!{\rm He}-\eta)_{\rm bound}\to ^3\!\!{\rm He}\, {\rm n}\, \pi^0$ process, 
 is at the level of the cross section values expected based on the hypothesis that the total cross section of the production of virtual $\eta$ meson just below the threshold is equal to the cross section of the production of the real $\eta$ meson above the threshold,  which is 
 about 15~nb for the ${\rm dd} \to^4\!\!{\rm He}\,\eta$~\cite{Budzanowski:2008qx,Wronska,Willis,Frascaria}. 
 In the ${\rm dd}\to (^4\!{\rm He}-\eta)_{\rm bound}\to \
 ^3\!{\rm He}\, {\rm p}\, \pi^-$ reaction, the more quantitative estimation of the cross section based on the approximation of the scattering amplitude for two body processes results in the value of 4.5~nb~\cite{Wycech-Krzemien}.
A much higher relative precision was achieved in the most recent high statistics search for the $^3$He-$\eta$ bound state where the limit of about $\sim$15~nb for the ${\rm pd} \to (^3\!{\rm He}-\eta)_{\rm bound}\to {\rm d p} \pi^0 \to {\rm d p}\, \gamma\gamma$ process (see right panel of Fig.~\ref{upper_limit_1d}) is more than an order of magnitude lower than the close-to-threshold total cross section for creation of the real $\eta$ meson in the ${\rm pd} \to ^3\!\!{\rm He}\,\eta$ process, which is
about 400~nb~(see Fig.~\ref{fig:3He-photo-totalcross}).

\begin{figure}[h]
\centering
\includegraphics[width=0.8\textwidth]{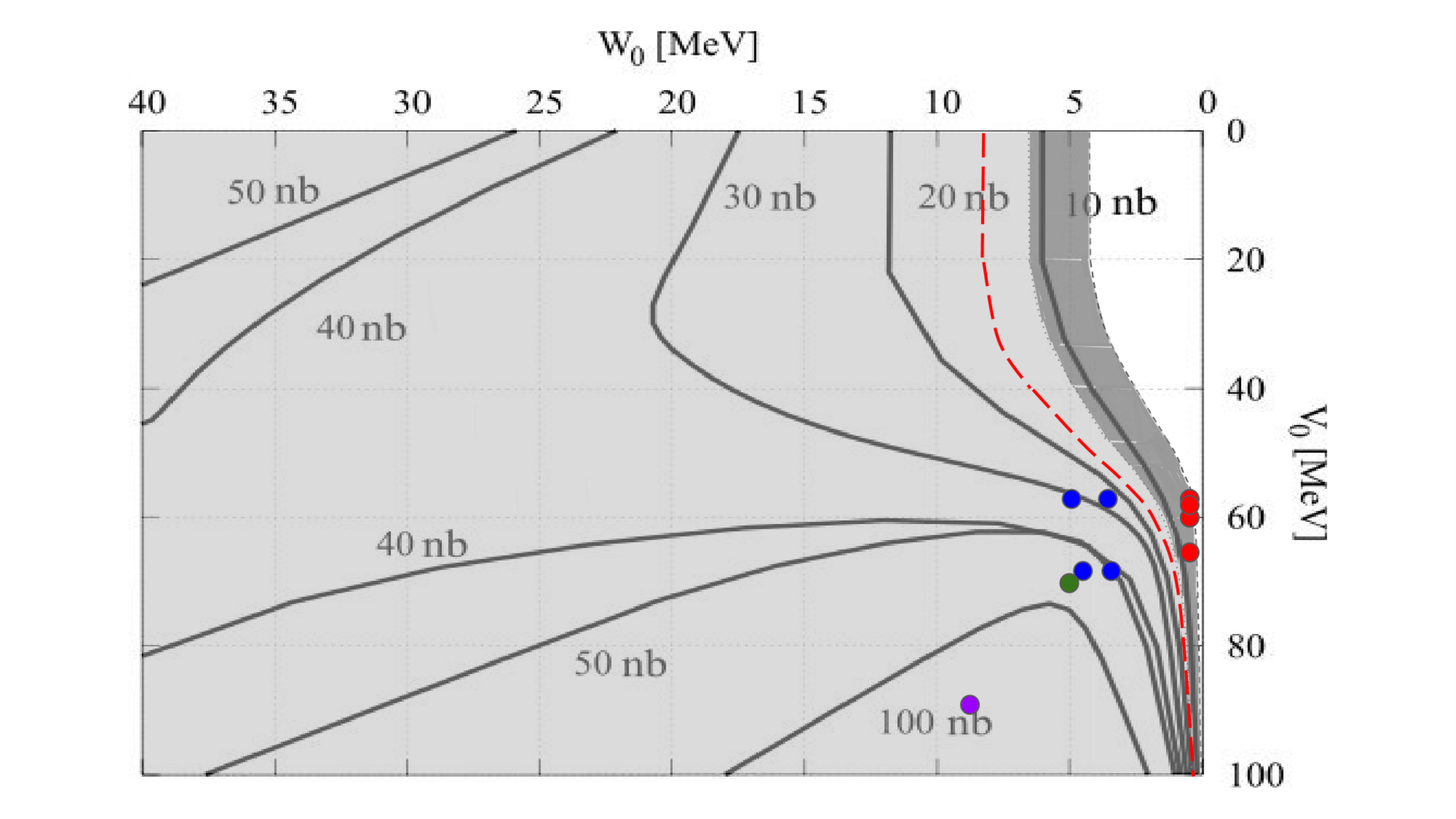}
\caption{ Contour plot in V$_0$-W$_0$ plane of the cross section for the $\eta$-mesic $^4$He production in dd reactions. The red curve separates the allowed parameter region (on the right) from the excluded region determined based on the  experimental limits of the cross sections. Dots correspond to the optical potential parameters of the
predicted $\eta$-mesic $^4$He states (see text for details). The figure is taken from~\cite{Skurzok:2018paa}.
~\label{exclussion-cross}}
\end{figure}

First quantitative estimations of the eta-mesic $^4$He production cross section as a function of the complex optical potential parameters ($V_0,W_0$) were presented in reference~\cite{Nikeno}.
In the model~\cite{Nikeno}, the Green’s function
technique is used to sum up all $\eta$-$^4$He final states for the estimation of the fusion and $\eta$ meson production processes.
The results compared to the experimentally determined excitation functions resulted in the determination of the exclusion region of ($V_0,W_0$) parameters~\cite{Magda-rev,Skurzok:2018paa} which is presented in Fig.~\ref{exclussion-cross}. The parameters resulting in a cross section larger than 10.7~nb (dark shaded are include systematic errors) were excluded at the 90\% CL~\cite{Skurzok:2018paa}. The figure indicates that 
most of the model parameter space is excluded, except for the
values of the real and imaginary parts of the
potential 
where $V_0$ is 
in the range 
$\sim$-60~MeV to 0 and 
$W_0$ is between
$\sim$-7~MeV and 0~\cite{Skurzok:2018paa}.  
Purple and green dots in the excluded region denote predictions based on the few body formalism with an optical model~\cite{Barnea:2017oyk} where the complex $\eta$-nucleon scattering amplitude is obtained (i) from a K-matrix description of the $\pi$N, $\pi \pi$N, $\eta$N and $\gamma$N coupled channels and fit to existing data~(purple dot)~\cite{GW} and (ii) a chirally motivated separable potential model with the parameters fitted to $\pi {\rm N} \rightarrow \pi {\rm N}$ and $\pi {\rm N} \rightarrow \eta {\rm N}$ data (green dot)~\cite{CS}. Blue dots indicate results obtained for 
a class of potentials including Gaussian, exponential and Hulthen~\cite{Liverts}. Red dots at the edge of the allowed parameter region present predictions of very narrow and weakly bound states of $^4$He-$\eta$, with binding energies and widths in the range of $\sim$2–230~keV and $\sim$8–64~keV respectively, that are found by solving the Klein Gordon equation as in~\cite{Nikeno}. These states correspond to the optical potential parameters $\vert V_0 \vert$ in the range from 58~MeV to 65~MeV and $W_0$~=~0.5~MeV.

\section{\textit{5. The $\eta'$ -nucleus interaction and the search for $\eta'$ mesic states}}

\subsection{5.1 The $\eta’$ -nucleus potential}

\subsubsection{5.1.1 Determination of the $\eta’$-proton scattering length}

Information on the $\eta'$-proton scattering length has been obtained from studies of the 
${\rm pp} \rightarrow {\rm pp} \eta'$ reaction near threshold at COSY \cite{Moskal:2000pu,Czerwinski:2014yot}. The first measurement showed that the $\eta'$-p scattering length is of the order of 0.1 fm. In the second measurement the reaction was studied with high statistics up to an excess energy of 11 MeV above threshold, where the cross section is clearly s-wave dominated. Fitting the excitation function, an analysis of the $\eta'$ final state interaction in $\eta'$ production in proton-proton collisions yields an $\eta'$-nucleon scattering length in free space of
\begin{equation}
a_{\eta' {\rm p}} = 
(0 \pm 0.43) + i (0.37^{~+0.40}_{~-0.16})~\mathrm{fm,}
\label{Eq:C11-a-eta-prime}
\end{equation}
indicating a relatively weak $\eta'$ nucleon interaction \cite{Czerwinski:2014yot}. 
Using Eqs.~(\ref{eq:eqa1},\ref{eq:eqa2},\ref{eq:eqa3}) the real and imaginary part of the $\eta'$-p scattering length can be converted into the real and imaginary part of the $\eta'$-nucleus potential
\begin{equation}
    U_0 = V_0+ i \cdot W_0 = -(0 \pm 37.9 + i \cdot 32.6^{~+35.2}_{~-14.1})~\mathrm{MeV}
\end{equation}
for comparison to direct determinations of the $\eta'$-nucleus potential parameters in the following sections.

Analyzing polarization observables and differential cross sections measured in the ${\rm \gamma p \rightarrow p \eta'}$ reaction
 \cite{GRAAL,CLAS,A2}, Anisovich {\it et al.} \cite{Anisovich:2018yoo} obtain within a coupled channels model a modulus of the $\eta'$p scattering length of  
\begin{equation}
|a_{\eta' {\rm N}}| = 0.403 \pm 0.015 \pm 0.060 ~\mathrm{fm}
\end{equation}
The phase has been determined to be 87$^{\circ} \pm 2^{\circ}$ which implies a small real part and sizable absorption, consistent with the result of \cite{Czerwinski:2014yot} but in conflict with the data on $\eta'$ 
photoproduction off nuclei discussed below. A purely imaginary $\eta'$ scattering length is however not expected as discussed in Section 3.1. Independent experiments and analyses are needed to clarify this point.

\subsubsection{5.1.2 Determination of the imaginary part of the $\eta'$ - nucleus potential from measurements of the transparency ratio}
\label{W_0}
The imaginary part $W(r)$ of the complex meson-nucleus potential $U(r) = V(r) + iW(r)$  is a measure for the absorption of the meson in the nuclear medium. Hereby $r$ is the distance from the centre of the nucleus which serves as a production target for the short-lived meson as well as an absorber. The reduction of the meson flux in the nuclear target can be quantified by the transparency ratio defined as \cite{Hernandez:1992rv,Cabrera:2003wb}
\begin{equation}
T_A=\frac{\sigma_A}{A\cdot \sigma_N} \label {TA}   
\end{equation}
which compares the meson production cross section off the nucleus with mass number A to A times the production cross section off the free nucleon. Thus, in case of no absorption, $T_A$ = 1, if secondary production processes can be neglected. Experimentally, the transparency ratio $T_A$ is determined by measuring the meson production cross section off the nucleus with mass number A and relating it to the production cross section off the proton or a light nucleus like $^{12}$C.

The absorption of the meson shortens its lifetime in the nuclear medium and therefore increases its width $\Gamma(\rho_N)$. For a linear density dependence the width $\Gamma(\rho_N)$ at nuclear density $\rho_N$ is given by
\begin{equation}
 \Gamma(\rho_N) =\Gamma_0 \cdot (\rho_N(r)/\rho_0) \label{Gamma}
\end{equation}
where $\Gamma_0$ is the width at normal nuclear density $\rho_0$
. The imaginary potential $W(r)$ is then related to the in-medium width $\Gamma_0$ via
\begin{equation}
 W(r) = -\frac{1}{2} \cdot \Gamma_0\cdot \label{width} \frac{\rho_N(r)}{\rho_0}
\end{equation}

\begin{figure}[b!]
\centering
{\includegraphics[width=0.36\textwidth]{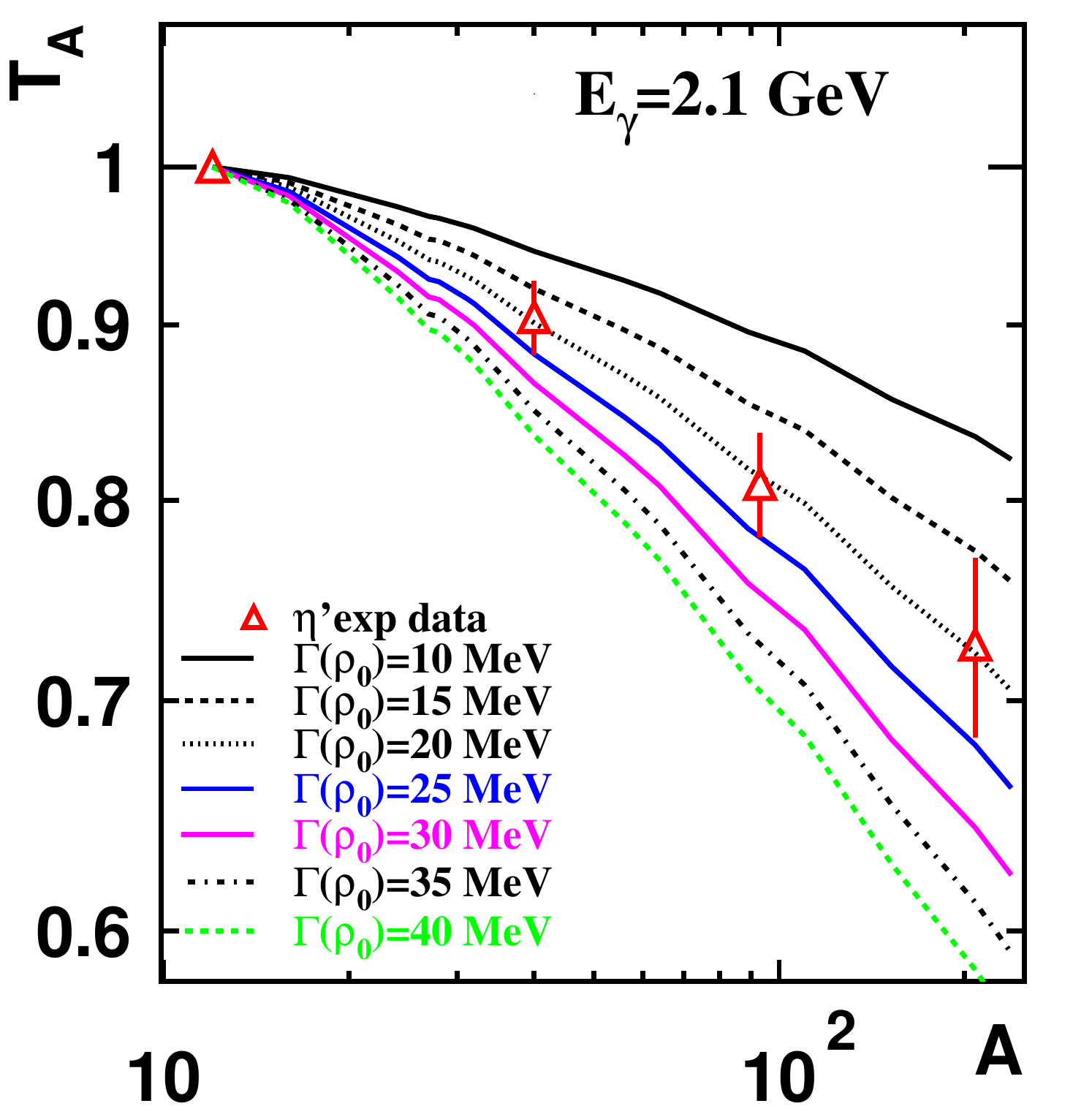}\includegraphics[width=0.64\textwidth]{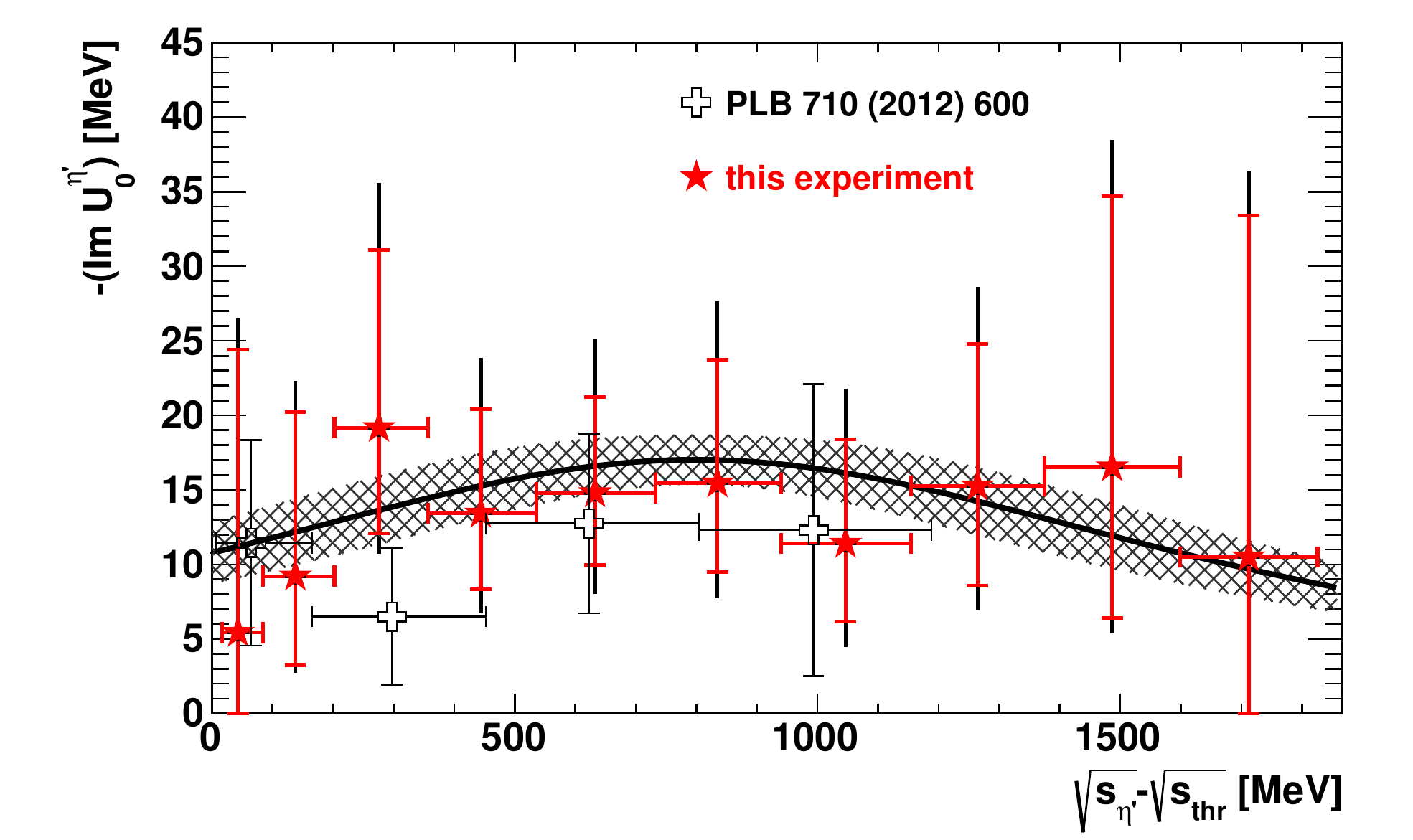}}
\caption{(Left): $\eta'$ transparency ratio for Ca, Nb and Pb normalized to that for C as function of the nuclear mass number in comparison to transport model calculations for different in-medium widths $\Gamma_0$ \cite{Nanova:2012vw}.(Right): Imaginary part of the $\eta'$-Nb potential as function of the available energy (red stars) \cite{Friedrich:2016cms} in comparison to earlier measurements (open crosses) \cite{Nanova:2012vw}. The solid curve is a fit to the data with a Breit-Wigner function. The shaded area indicates a confidence level of $\pm 1 \sigma$ of the fit curve, taking statistical and systematic errors into account. The figures are taken from \cite{Nanova:2012vw,Friedrich:2016cms}, respectively. With kind permission of The European Physical Journal (EPJ).}
\label{fig:Gamma_eta'}
\end{figure}

As described in \cite{Metag:2017yuh,Nanova:2012vw,Friedrich:2016cms} the transparency ratio can be calculated with  Glauber-, transport- or collisional-model approaches for any in-medium width $\Gamma_0$. Conversely, comparing the measured transparency ratio to the model results, the in-medium width $\Gamma_0$ and thus by Eq.(\ref{width}) the imaginary part of the meson-nucleus potential can be deduced. 

The result of the first measurement of the $\eta'$ transparency ratio in the $\gamma A \rightarrow \eta' + X$ reaction for Ca, Nb, and Pb is shown in Fig.\ref{fig:Gamma_eta'} (left). A comparison of the experimental results with transport calculations leads to an in-medium width at normal nuclear density of about 20 MeV \cite{Nanova:2012vw}.

For the existence and observability of meson-nucleus bound states the in-medium width near the production threshold, i.e. at low meson momenta relative to the nuclear environment is decisive. For extrapolating to low momenta the transparency ratio has to be measured over a broad momentum or energy range range. As an example Fig.\ref{fig:Gamma_eta'} (right) \cite{Friedrich:2016cms} shows the imaginary potential for $\eta'$ mesons in Nb as function of the excess energy above threshold, derived from a Glauber model analysis of transparency ratio measurements for each excess energy bin. The extrapolation to the production threshold using different fit functions yields an imaginary part of the $\eta'$-Nb potential at normal nuclear matter density of $W(\rho_0)=(13\pm3({\rm stat.})\pm 3({\rm syst.}))$ MeV, consistent with the earlier result \cite{Nanova:2012vw}. Further details of this extraction and a discussion of uncertainties can be found in the original literature \cite{Friedrich:2016cms} or the review \cite{Metag:2017yuh}.

\subsubsection{5.1.3 Determination of the real part of the $\eta'$ nucleus potential by measuring excitation functions and/ or momentum distributions} 
\label{V_0}
A measurement of the meson production cross section as a function of the incident beam energy is sensitive to in-medium modifications of the meson since a downward mass shift would lower the production threshold and thus increase the the production cross section at a given incident beam energy due to the enlarged phase-space. Furthermore, transport calculations \cite{Weil:2012qh} have demonstrated that also the momentum distribution of mesons produced off nuclei is sensitive to the in-medium properties of the meson. When leaving the nucleus a meson with reduced in-medium mass has to get back to its free vacuum mass. The missing mass has to be generated at the expense of its
kinetic energy. Consequently, this energy-to-mass conversion shifts the meson momentum distribution to lower average values.
\begin{figure}[h!]
\centering
{\includegraphics[width=0.50\textwidth]{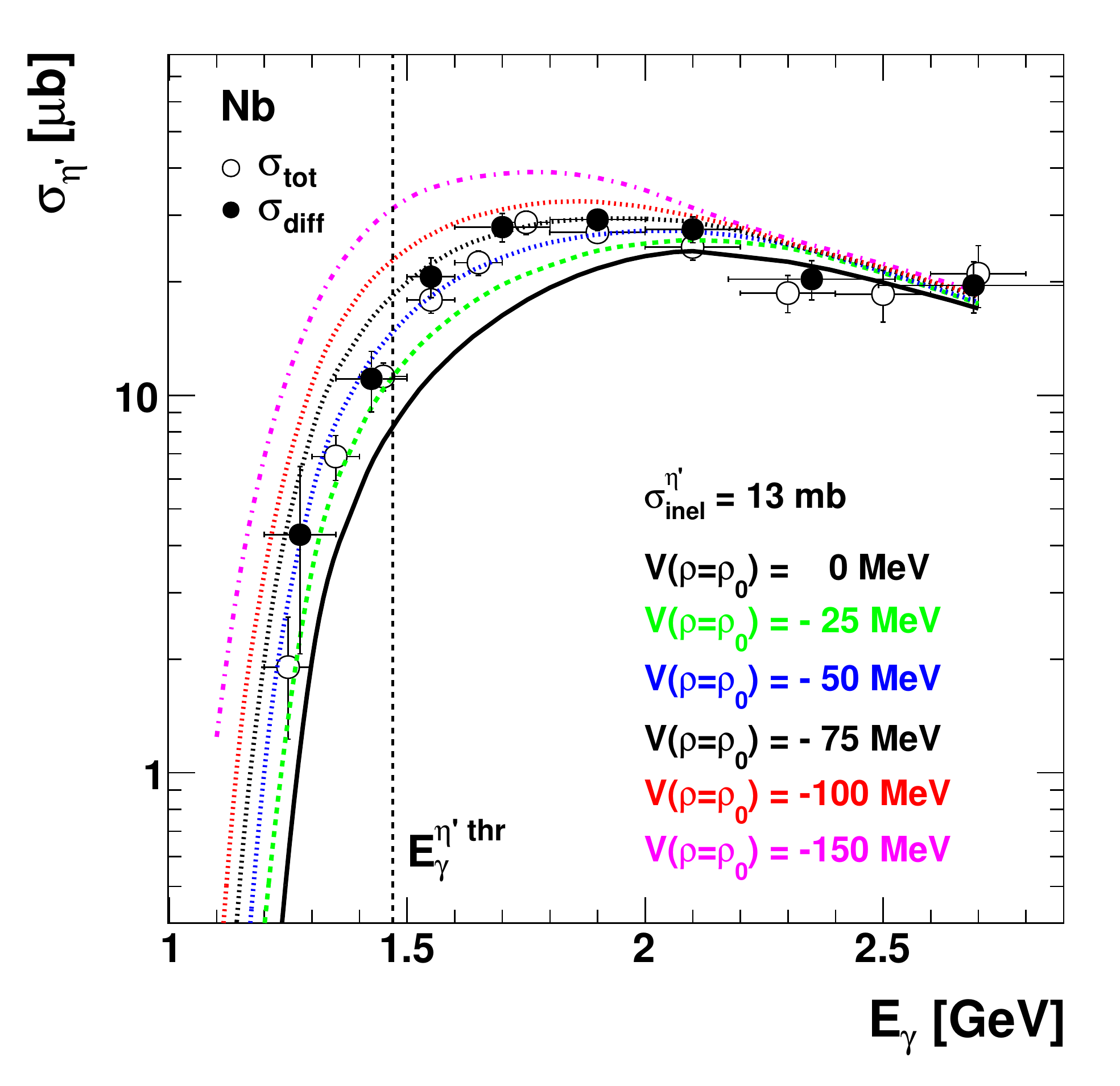}\includegraphics[width=0.51\textwidth]{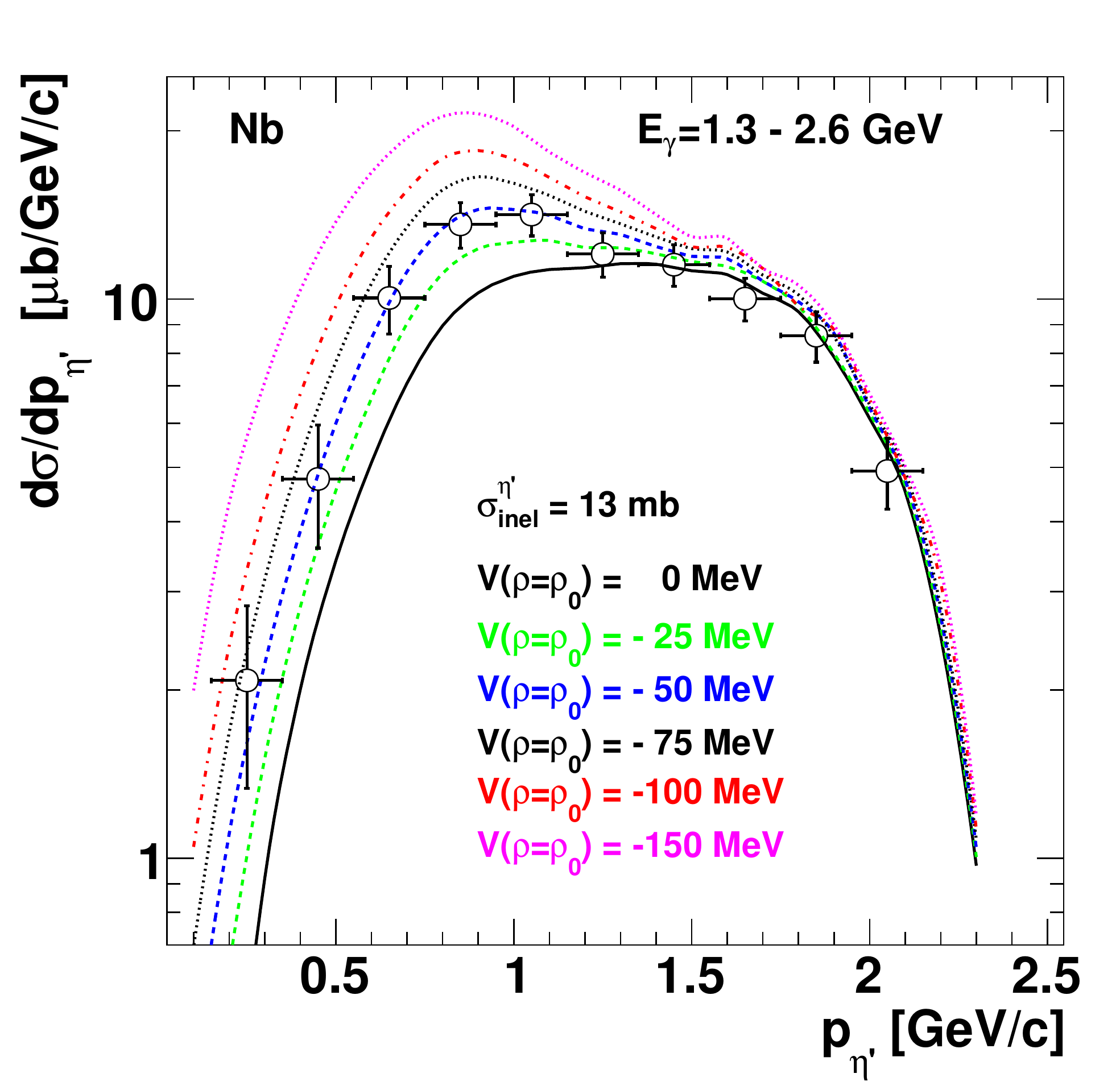}}
\caption{Excitation function (left) and momentum distribution (right) of $\eta'$ mesons produced in the reaction $\gamma {\rm p} \rightarrow \eta' {\rm X}$ in comparison to collision model calculations for different values of the real part of the $\eta'$-Nb potential. The figures are taken from  \cite{Nanova:2016cyn}.}
\label{fig:eta' mass}
\end{figure}

\begin{figure}[h!]
\centering
\includegraphics[width=0.8\textwidth]{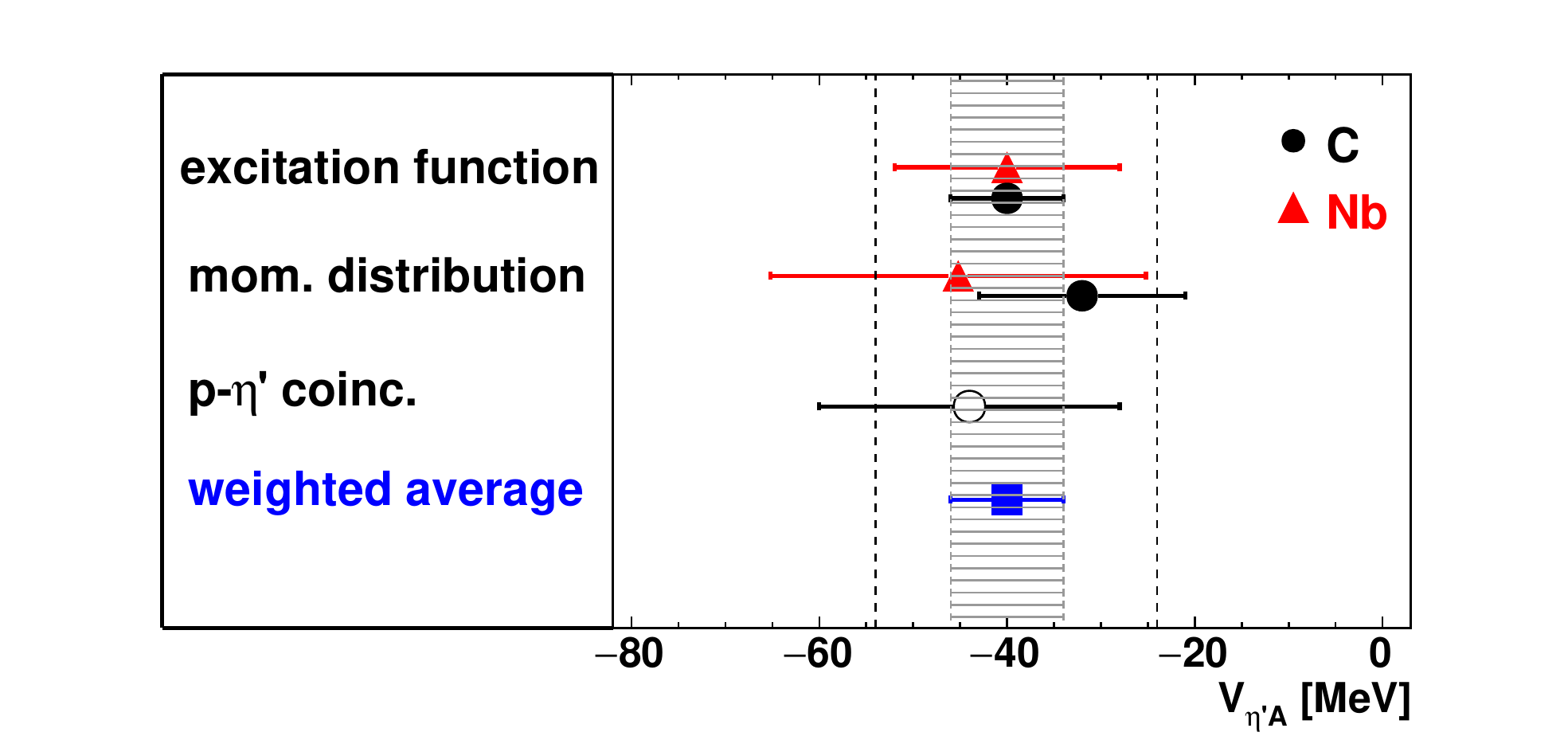}
\caption{Depth of the real part of the $\eta'$-nucleus potential derived from inclusive measurements on C \cite{Nanova:2013fxl} and Nb \cite{Nanova:2016cyn} and from an $\eta'$-p coincidence experiment \cite{Nanova:2018zlz}. The weighted overall average is represented by a blue box. The shaded area indicates the statistical error. The dashed lines mark the range of systematic uncertainties.The figure is taken from
\cite{Nanova:2018zlz}. With kind permission of The European Physical Journal (EPJ).}
\label{fig:V0}
\end{figure}

The measurement of the $\eta'$ excitation function and momentum distribution have both been used to extract the in-medium mass shift of the $\eta'$ meson in Nb as shown in Fig.\ref{fig:eta' mass}. An enhancement of the total cross section and a shift towards lower momenta compared to a scenario without mass modification is observed. A quantitative  comparison with collision model calculations yields in-medium mass shifts of -(40$\pm$12) MeV and -(45$\pm$20) MeV, respectively. Further inclusive measurements have been performed on C \cite{Nanova:2013fxl} as well as a semi-inclusive study of low momentum $\eta'$ mesons in coincidence with high energy forward going protons which take over most of the momentum of the incident beam \cite{Nanova:2018zlz}. The values extracted for the real part of the $\eta'$-nucleus potential do not show a significant mass dependence and are summarized in Fig.\ref{fig:V0} \cite{Nanova:2018zlz}. 

\subsubsection{5.1.3 Parameters of the $\eta'$-nucleus potential}
As described in the preceding sections, the parameters of the $\eta'$-nucleus potential have not been directly measured but have been extracted from experimental observables such as transparency ratios, excitation functions and meson momentum distributions using transport and collision models and Glauber calculations. Only models have been used that have been widely tested and successfully applied in other areas of nuclear and hadron physics, giving consistent results in the present analysis. The experimental results have been reproduced in a long series of independent experiments over several years. In view of these consistencies the following final values of the real and imaginary part of the $\eta'$-nucleus potential are quoted:
 \begin{equation}
  V (\rho=\rho_0) = -(40 \pm 6 ({\rm stat}) \pm 15({\rm syst})) {\rm MeV}   
 \label{eqe4}
 \end{equation}
\begin{equation}
 W (\rho=\rho_0) = -(13 \pm 3 ({\rm stat}) \pm 3({\rm syst})) {\rm MeV} \label{eqe5}     
\end{equation}    
\noindent
Theoretical predictions for the real part of the $\eta$ - nucleus potential cover a broad range from -150 MeV \cite{Nagahiro:2006dr}, -80 MeV\cite{Sakai:2013nba}, -40 MeV\cite{Bass:2005hn} to 0 MeV \cite{Bernard:1987sx}. Only the result of \cite{Bass:2005hn} is close to the experimental value. A more detailed discussion of the experimental uncertainties and the comparison to theory is given in \cite{Metag:2017yuh}.

More direct information on the $\eta'$-nucleus potential will be accessible from the observation of $\eta'$-nucleus bound states. As the modulus of the real part of the potential is found to be about 3 times larger than the modulus of the imaginary part - a favourable condition for the observation of meson-nucleus bound states - the $\eta'$ meson appears to be a promising candidate in the search of mesic states.

\subsection{5.2 Direct searches for $\eta'$ mesic states}
\subsection{5.2.1 Search for $\eta'$-mesic states in the $^{12}{\rm C(p,d)}^{11} {\rm C} \otimes \eta'$ reaction}

The first pioneering experiment searching for $\eta^\prime$ bound states was performed in 2014 at the Fragment Separator (FRS) at GSI using the $^{12}$C(p,d) reaction \cite{Tanaka:2016bcp,Tanaka:2017cme}. The incident proton energy of 2.5 GeV was chosen to achieve almost recoil-less production of the $\eta'$ meson. A deuteron with high momentum is ejected to forward angles, while the $\eta^\prime$ meson produced with low momentum could be bound to the $^{11}$C. In the experiment only the deuteron momentum distribution has been measured, applying missing mass spectrometry. 
\begin{figure}[h!]
\centering
{\includegraphics[width=0.45\textwidth]{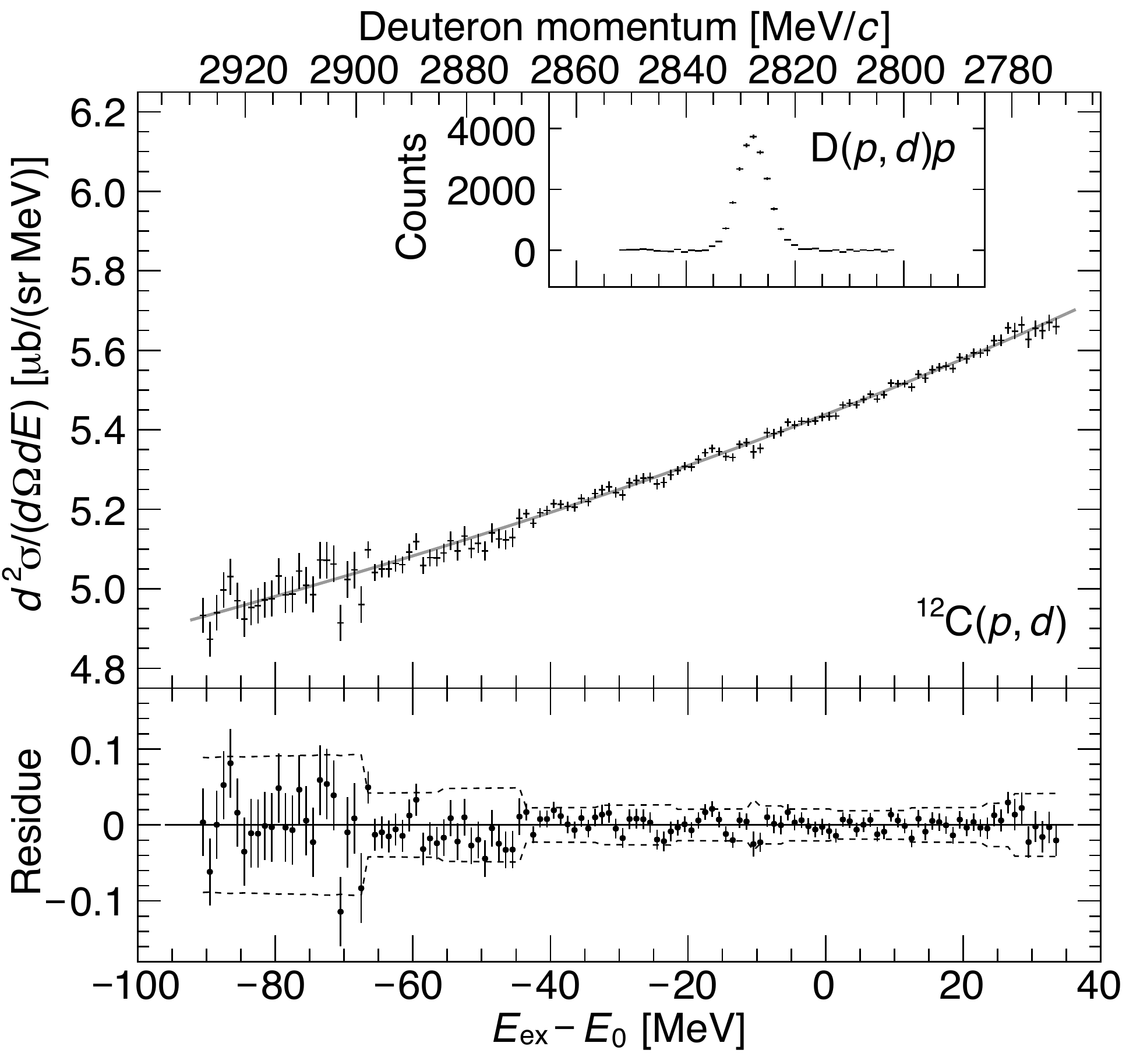}\includegraphics[width=0.56\textwidth]{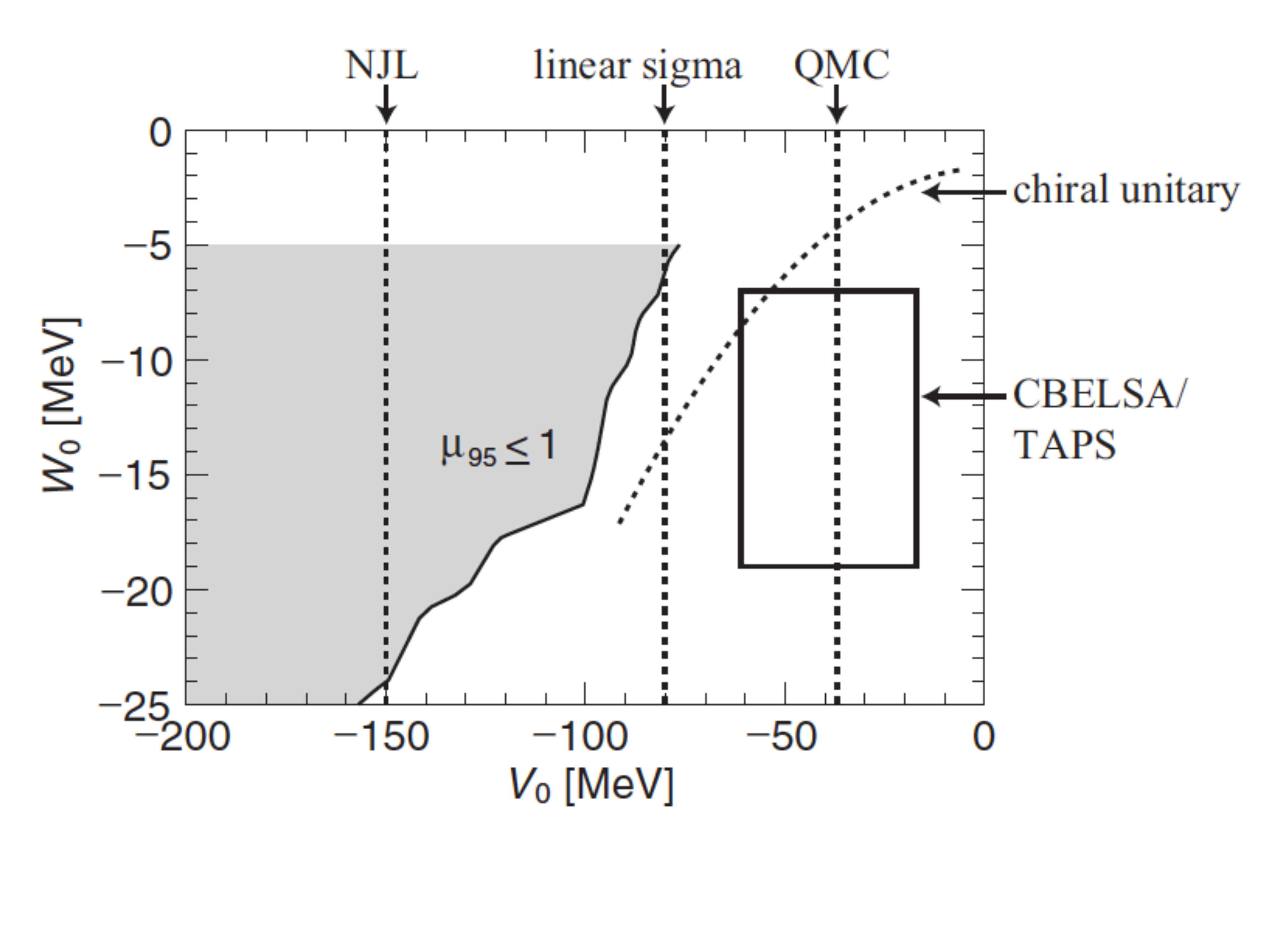}}
\caption{(Left):Excitation energy spectrum of $^{11}$C measured in the $^{12}$C(p,d) reaction at a proton energy of 2.5 GeV. The abscissa represents the excitation relative to the $\eta'$ production threshold E$_0$ = 957.78 MeV. The grey curve is a third-order polynomial fit to the data.The inset shows the momentum spectrum measured in the elastic D(p,d)p reaction. (Bottom panel): Fit residues with 2$\sigma$ envelopes. (Right): the imaginary part versus the real part of the $\eta'$-nucleus potential. The shaded area above the solid curve indicates the parameter range which is excluded by the FRS measurement \cite{Tanaka:2016bcp,Tanaka:2017cme}. Summing systematic and statistical errors the box represents the range of V$_0$ and W$_0$ parameters determined by the CBELSA/TAPS collaboration \cite{Nanova:2013fxl,Nanova:2016cyn,Friedrich:2016cms,Nanova:2018zlz} (s. section 5.1).Theoretical predictions with the NJL model \cite{Nagahiro:2006dr}, the linear sigma model \cite{Sakai:2013nba}, the QMC model \cite{Bass:2005hn} and the chiral unitary approach \cite{Nagahiro:2011fi} are shown by dashed lines.The figures are taken from \cite{Tanaka:2016bcp,Tanaka:2017cme}, respectively}
\label{fig:Tanaka}
\end{figure}

In Fig.~\ref{fig:Tanaka} (Left) the measured excitation spectrum of the $^{12}$C(p,d) reaction near the $\eta^\prime$ emission threshold is shown \cite{Tanaka:2016bcp}. Because of the multi-pion background no narrow structure has been observed in spite of the extremely good statistical sensitivity. An upper limit for the formation cross section of $\eta'$-mesic nuclei of $\approx$20 nb/(sr MeV) near the threshold has been deduced. In a detailed analysis the experimental spectrum has been compared to theoretical predictions \cite{Nagahiro:2012aq} for different potential parameters ($V_{0}, W_{0}$), allowing an exclusion of certain parameter ranges as indicated in Fig.~\ref{fig:Tanaka} (right). 
A strongly attractive potential of $V_{0}\approx$-150 MeV predicted by the NJL model calculations~\cite{Nagahiro:2012aq} can be rejected. Other sets of predicted potential parameters, also shown in Fig.~\ref{fig:Tanaka} (right), can not be excluded. The experimental result is consistent with the potential parameters determined in the photoproduction experiments discussed in section 5.1. 

An improved follow-up experiment \cite{Itahashi}  is in preparation providing higher sensitivity by combining missing mass spectrometry with simultaneous detection of protons from the decay of the $\eta^\prime$-mesic states. An important decay mode is two-nucleon absorption 
$\eta' {\rm N N} \rightarrow 
{\rm N N}$ yielding protons with 300 - 600 MeV in the laboratory. Simulations have shown that by selecting energetic protons in the backward angular range the multi-pion background can be efficiently suppressed. In comparison to the pioneering experiment the signal-to-background ratio will be improved by two orders of magnitude. The experiment is scheduled for 2022.

 \subsubsection{5.2.2 Search for $\eta'$-mesic states in the $^{12}{\rm C}(\gamma,{\rm p})$ reaction}
 
 The idea of combining missing mass spectrometry with coinicident detection of decay products of 
the $\eta'$-mesic state has already been realized in the experiment by the LEPS2/BGOegg collaboration at Spring-8 \cite{Tomida:2020yin}.

Using photon beams of 1.3-2.4 GeV generated by laser backscattering the following reaction has been studied in small momentum transfer kinematics:
\begin{equation}
 \gamma +^{12}{\rm C} \rightarrow {\rm p}_f + \eta' \otimes ^{11}{\rm B}   
\end{equation}
in coincidence with $\eta$-proton pairs from
\begin{equation} 
\eta' + {\rm p} \rightarrow \eta +{\rm p}_s
\end{equation}
which is expected to be the strongest
absorption process for an $\eta'$ meson bound to a nucleus. The forward going proton p$_f$ is used for missing mass spectrometry while the sideward going proton p$_s$ together with the $\eta$ meson tags the decay of the $\eta'$ mesic state. Since the bound $\eta'$ meson is almost at rest the $\eta$ and proton will be emitted nearly back-to-back in the laboratory. By the simultaneous measurement of the $(\eta, {\rm p}_s)$ pair and the forward going proton p$_f$ the multi-pion background can be effectively suppressed. 

\begin{figure}[h!]
\centering
\includegraphics[width=0.8\textwidth]{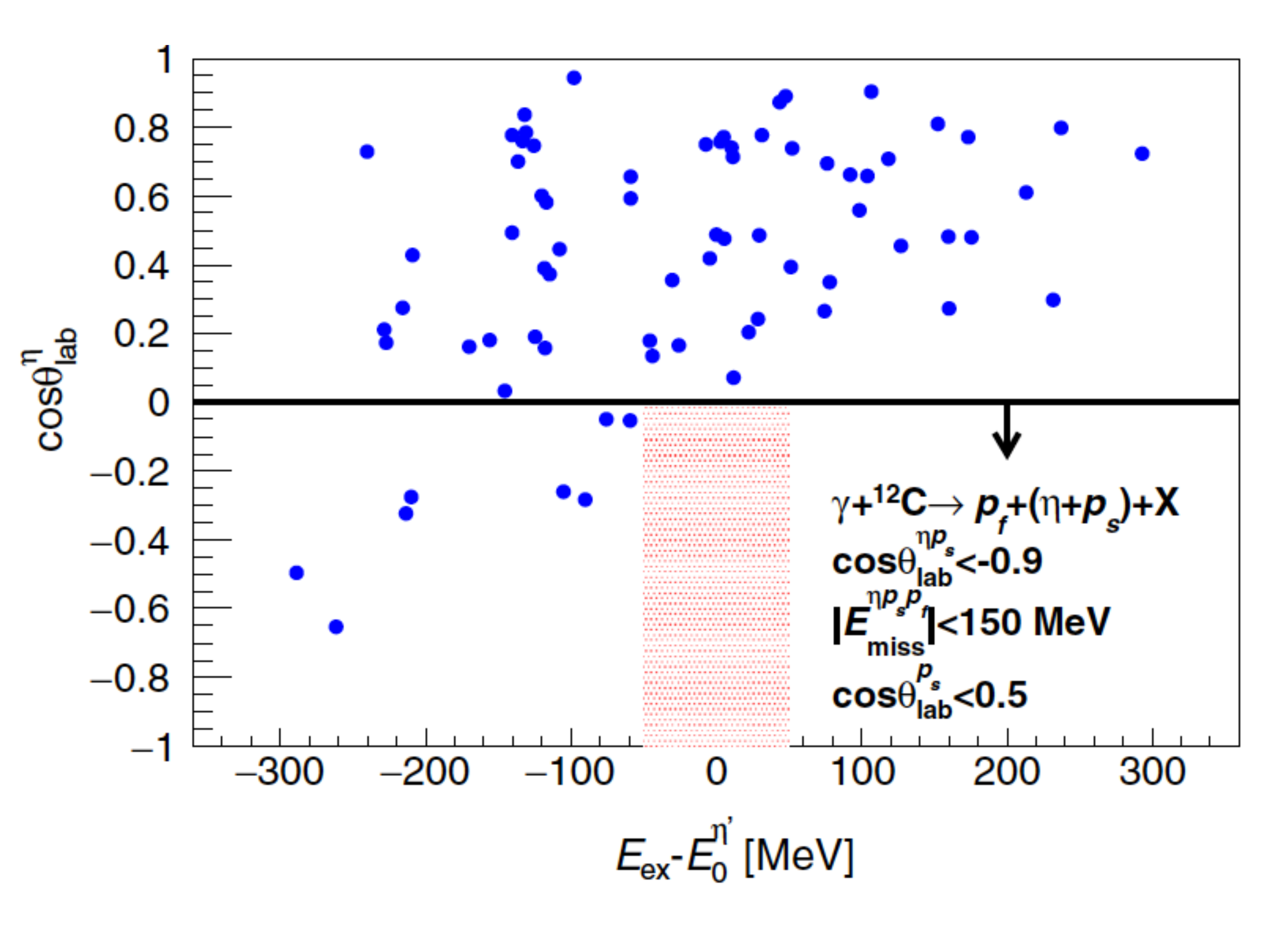}
\vspace{-0.5cm}
\caption{Two dimensional plot of  $\cos \theta_{\rm lab}^{\eta}$ vs. $E_{ex} - E_0^{\eta'}$ for the $(\eta+{\rm p}_s),{\rm p}_f$ coincidence data. The expected signal region is marked by red hatching. The figure is taken from \cite{Tomida:2020yin}. With kind permission of the American Physical Society.}
\label{fig:Tomida}
\end{figure}
As shown in Fig.~\ref{fig:Tomida} no signal events have, however, been observed in the bound-state region leading to an upper limit of the signal cross section of 2.2 nb/sr at the 90$\%$ confidence level for opening angles $\cos \theta_{\rm lab}^{{\rm p}\eta} \ge -0.9$. An attempt has been made to extract from this result an upper limit for the formation cross section of an 
$\eta'\otimes ^{11}$B state with subsequent ($\eta+{\rm p}_s)$ decay. By comparing to a theoretical cross section \cite{Nagahiro:2017xpi} the real part $V_0$ of the  $\eta'-^{11}$B optical potential and the branching fraction BR$_{\eta' {\rm N} \rightarrow \eta {\rm N}}$ for the decay via the ($\eta+{\rm p}_s$) channel have been constraint. Hereby, the theoretical cross section has been normalized to reproduce the measured $\eta'$ cross section in the unbound region near the production threshold. Tomida et al. \cite{Tomida:2020yin} deduce an upper limit for the branching ratio BR$_{\eta' {\rm N} \rightarrow \eta {\rm N}}$ of 24$\%$ for a potential depth $V_0$ = -100 MeV and of 80$\%$ for $V_0$ = -20 MeV, respectively, at the 90$\%$ confidence level. They conclude that BR$_{\eta' {\rm N} \rightarrow \eta {\rm N}}$  is small and/or the real part $V_0$ of the $\eta'$-nucleus potential is shallow. However, in a comment to this work, Fujioka et al. \cite{Fujioka:2021ewc} point out several uncertainties in this analysis and claim that $V_0$ and BR$_{\eta' {\rm N} \rightarrow \eta {\rm N}}$ may be over constrained.

\section{\textit{6. Conclusions}}
In spite of numerous experimental efforts, $\eta$- and $\eta‘$ – nucleus bound states have so far not been directly observed. However, information on the strength of the $\eta$ and $\eta’$ interaction with nuclei has been deduced. A strong $\eta$ - nucleus interaction has been experimentally established in hadron- and photon- induced reactions independent of the entrance channel dynamics. Comparing model calculations with cross section measurements as a function of the excitation energy, parameters of the real and imaginary part of the optical potential have been constrained to $V_0 = -60$ to 0 MeV and $W_0 = -5$ to 0 MeV in the case of $^4$He. For the 
$\eta’$ meson the corresponding potential parameters are in the range of  $V_0 \approx $ -40 MeV and $W_0 \approx$ -13 MeV. Both parameter sets indicate an attractive meson-nucleus interaction and a relatively weak meson absorption.

On theoretical side, interest is in understanding the 
role of gluonic degrees of freedom in the QCD phase diagram and (partial) restoration of axial U(1) symmetry at finite nuclear density.
What makes the $\eta'$ and $\eta$ special compared to other mesons not sensitive to the gluonic potential that gives the
$\eta'$ and $\eta$ extra mass.
Future work might focus on extending model treatments of the $\eta'$ and $\eta$ in medium 
to make closer, more direct, connection with QCD and the symmetries of anomalous Ward identities
at
finite density 
as well as at finite temperature.

The search for $\eta$, $\eta’$ mesic states continues. Promising results on the possible existence of a (virtual) $\eta$-d state have recently been presented by the ELPH group and need further independent confirmation. The coherent $\pi^0\eta$ photoproduction off nuclei appears to be a promising approach for further searches of $\eta$-nucleus bound states. Selecting events with a high $\pi^0$-momentum will render $\eta$-mesons with a small momentum relative to the intact nucleus, allowing for an enhanced formation of a bound state. Measuring simultaneously the production and decay of the mesic state, the approach pioneered by the Spring 8 experiment, the planned WASA@FRS measurement will hopefully reach the required sensitivity for the observation of an $\eta’$-nucleus bound state. The sensitivity of the Spring 8 experiment will be increased by studying a variety of other potential decay modes of the $\eta'$-nucleus bound states. New results are expected in the coming years.

\section*{Acknowledgements}
P.M. acknowledges the support from the Polish National Science Center through grant No. 2016/23/B/ST2/00784 and the EU Horizon 2020 research and innovation programme, STRONG-2020
project, under grant agreement No 824093.

%
%
%
%

\end{document}